\documentclass{appolb}
\usepackage{graphics}
\usepackage{epsfig}
\usepackage{esvect}
\usepackage{epstopdf}
\usepackage{amssymb}
\usepackage{amsbsy}

\begin{document}

\title{Ground-state nuclear properties of neutron-rich copper isotopes and lepton capture rates in stellar matter%
}
\author{Jameel-Un Nabi $^{1}$ \footnote{Corresponding author}, Tuncay Bayram $^{2}$ and Muhammad Majid
$^{1}$
\address{$^{1}$Faculty of Engineering Sciences, GIK
Institute of Engineering Sciences and Technology, Topi 23640, Khyber
Pakhtunkhwa, Pakistan\\
$^{2}$Department of Physics, Faculty of Science, Karadeniz Technical
University, Trabzon, Turkey.}\\
%
\address{{$^{*}$jameel@giki.edu.pk, t.bayram@ktu.edu.tr, majid.phys@gmail.com}}
}
\maketitle
%

\begin{abstract}
This study essentially consists of two separate investigations where
we study neutron-rich isotopes of copper. Two different nuclear
models were selected to perform the two investigations. In the first
part of this paper nuclear \textit{ground-state} properties of
neutron-rich copper isotopes (72 $\leq$ A $\leq$ 82) have been
studied with the help of  the relativistic mean field (RMF) model.
Using quadrupole moment constrained RMF calculation with DD-ME2 and
DD-PC1 density dependent interactions, the ground-state binding
energies, charge radii, proton and neutron radii, quadrupole moments
and deformations of $^{71-82}$Cu were calculated. The results are in
decent agreement with the limited number of available measured data
and previous theoretical calculations. Additionally, we calculate
the potential energy curves for these copper isotopes and discuss their ground-state geometrical shapes.\\
The second half of this paper is dedicated to calculation of lepton
capture rates in stellar environment. Recently allowed Gamow-Teller
(GT) and unique first-forbidden (U1F) $\beta$-decay rates of the
selected neutron-rich Cu nuclide under stellar environment were
presented. However the lepton capture rates were not calculated.
They were recently calculated and reported in this paper. Ground and
excited states GT and U1F strength functions were calculated in a
microscopic way employing the deformed proton-neutron quasiparticle
random phase approximation (pn-QRPA) model with schematic separable
GT forces. The deformed pn-QRPA model satisfied the model
independent Ikeda sum rule. The lepton capture rates were computed
on a wide temperature range of (0.01--30)$\times10^{9}$K and stellar
density range of (10--10$^{11}$) g/cm$^{3}$. We compared our
computed half-lives (GT + U1F) with previous theoretical and
measured results. Our calculated terrestrial half-lives agree well
with the measured ones. We also present the percentage contribution
of positron capture  to $\beta$-decay rates under stellar
conditions. Our study shows that, at high stellar temperatures,
allowed GT and, specially, U1F positron capture rates dominate the
competing $\beta$-decay rates. For a better description of
presupernova evolutionary phases of massive stars, simulators are
recommended to take into account lepton capture rates on
neutron-rich copper isotopes presented in this work.

\end{abstract}

\PACS{21.60.Jz, 23.40.-s, 26.50.+x, 97.10.Cv}

\vspace{2pc} \noindent{\it Keywords}: Relativistic mean field model;
Deformed pn-QRPA model; Potential energy curves; Ground-state
nuclear properties; Gamow-Teller transitions; unique first-forbidden
transitions; electron and positron capture rates.
%
%

\section{Introduction}
It is desirable to use a single nuclear model for prediction of
various nuclear ground-state properties over an isotopic chain.
Various nuclear models have been employed in the past to do the
needful and may broadly be classified into three main groups namely
macroscopic models, macroscopic-microscopic models and microscopic
models. An example from the macroscopic models would include the
Bethe-Weizs\"{a}cker mass formula \cite{Wei35} . FRDM (Finite Range
Droplet Model)~\cite{moller1997} is a good example from
macroscopic-microscopic models. The HF (Hartree-Fock)
method~\cite{Flo73} and the RMF (relativistic mean field) model
\cite{Walecka74} may be cited as examples belonging to the
microscopic genre. Every model has associated pros and cons.
Self-consistent calculation using mean field approximation with
effective interactions are suitable alternate approaches for a
better prediction of nuclear properties of finite nuclei. Recently
the RMF model was employed to calculate ground-state energies,
deformations and sizes of 1897 even-even nuclei, for atomic number
between 10 and 110, using the nonlinear RMF force NL3*
\cite{bayram2013a}. More recently, a great effort had been done for
developing RMF mass model with density-dependent meson coupling
interactions comparable to the most accurate non-relativistic
microscopic ones~\cite{arteaga2016}.

In the first investigation of our study, we employ the axially
deformed RMF model with density-dependent functionals to study
ground-state nuclear properties of copper (Cu) (neutron-rich)
isotopes. Isotopes of Cu are believed to play a key role in the
presupernova evolution of massive stars \cite{auf94,heg01}. The RMF
model is successful in determining the ground-state nuclear
properties  close to neutron and proton drip lines. The RMF model,
with adjustable small number of parameters, can provide correct
predictions of various ground-state properties of isotopes not only
along the stability line but also far from it
\cite{bayram2013a,Gambhir90,Ring96,Vretenar05}. For reliable
determination of ground-state properties of copper isotopes,
quadrupole moment constrained calculation has been carried out in
this model. Potential energy curves (PECs) for $^{71-82}$Cu,
according to the quadrupole deformation parameter ($\beta_{2}$),
have been obtained using the RMF model. Lowest binding energy ($BE$)
of PEC for each nuclei was taken as the ground-state $BE$. The
ground-state shape evolution of $^{71-82}$Cu nuclei is discussed by
using the PEC results. Calculated values of $BE$ for copper isotopes
were compared with the measured values and calculated results of the
FRDM model in the present work. Proton, neutron and charge radii of
a nucleus are important nuclear properties and they are directly
related with the size of nuclei. The quadrupole moment of nuclei,
which is related to the deformation, is also one of the crucial
nuclear properties. In this investigation we calculate the binding
energy, proton and neutron radii, root mean square (rms) charge
radius, deformation parameter and quadrupole moment of neutron-rich
copper isotopes ($^{71-82}$Cu), bearing astrophysical importance,
using the RMF model.

Many efforts have been done in recent years to investigate the
nuclear masses and charge-changing transition rates of
neutron-abundant species at radioactive ion-beam facilities. However
in terrestrial laboratories many of the exotic nuclear species
cannot be studied and  problem at hand is best resolved by
considering theoretical approaches. Study of evolution process of
massive stars and related nucleosynthesis mechanism have attracted
astrophysicists for many decades. An iron core is left behind during
the last stage of star burning. Photodisintegration of iron and
capturing of free electrons make the core unstable and ensue the
collapse. Collapse process is very sensitive to the
electron-to-baryon ratio (Y$_{e}$) and entropy of the stellar core
\cite{bet79}. The weak-rates, consisting of the lepton capture and
emission rates, play a crucial role in controlling these parameters.
The simulation of core collapse of massive stars depends heavily on
capturing of the electrons \cite{hix03}. Due to capturing of
electrons, the Y$_{e}$ content is effectively decreased during the
initial stages of collapse. Capturing of electrons decreases the
number of electrons responsible for degeneracy pressure, while
effect of positron capture (PC) acts in the opposite direction. Both
processes can lead to generation of (anti)neutrinos which in turn
lead to streaming out energy and entropy from the stellar core for
density range of $\rho\leq$ 10$^{11}$ gcm$^{-3}$. The importance of
electron capture (EC) during the presupernova evolutionary phases of
massive stars may be seen from Ref. \cite{bet90}. The significance
of PC is pivotal in the stellar core, particularly at high core
temperatures and low neutron density regions. During such
situations, a slightly larger concentration of e$^{+}$ particles can
be obtained from the stable conditions of $\gamma + \gamma$
$\longleftrightarrow$ e$^{-}$ + e$^{+}$ which favors the
electron-positron pairs. The chance for the occurrence of
equilibrium and the competition between PC on neutrons and EC on
protons are considered as the central and decisive constituents for
modeling of Type-II supernovae mechanism (for a detailed discussion
see \cite{nab07}).

At low temperature range ($\sim$ 300-800 keV) and intermediate
stellar densities ( $\leq$ 10$^{10}$ gcm$^{-3}$), EC generally
occurs on nuclei in the vicinity of A $\sim$ 60 \cite{jan07}. As the
chemical potential for electrons and the Q-value approach each
other, the corresponding stellar weak-rates become much responsive
to the detailed analysis of the Gamow-Teller (GT) strength
distributions. However, for bigger values of density and
temperature, nuclide, in the mass range  A $>$ 65, become more
abundant, and the chemical potential for the electrons are
appreciably greater than the Q-value. In this scenario GT centroid
energy and total GT strength are pre-requisites for the
determination of EC rates. At higher stellar density ($>$ 10$^{10}$
gcm$^{-3}$), the electron chemical potential becomes greater than 2
MeV. Under prevailing conditions forbidden transitions along with
allowed GT transitions need to be taken into account for a better
description of the presupernova structure. Detail discussion can be
seen in Refs. ~\cite{jan07} and \cite{lan03}.

The first-forbidden (FF) transitions may contribute in lowering the
half-lives for neutron-rich isotopes. In this regard a first attempt
was made by Homma et al. \cite{Hom96} to quantify the contribution
of FF transitions to $\beta$-decay half-lives (T$_{1/2}$) by
employing the proton-neutron quasiparticle random phase
approximation (pn-QRPA) model. The QRPA + gross theory framework
\cite{Mol03} also considered the effects of FF transitions. More
recently, self-consistent density-functional + continuum QRPA
Ref.~\cite{bor06} approach, extended by Borzov, reported the allowed
and FF half-lives for the $r$-process simulations. Borzov study
suggests that FF transitions contribute a small appreciable
correction to the $N=50$ and 82 isotones, but significantly reduce
the T$_{1/2}$ values near the $N=126$ isotones. The same FF
contribution to total $\beta$-decay half-lives were also studied by
\cite{Zhi13} using large scale shell model approach. Recently the
deformed pn-QRPA model was used for the calculation of both GT and
U1F (having $|\Delta$J$|$ = 2) transitions of $^{72-82}$Cu nuclide
in stellar scenario \cite{Maj17}. There the authors concluded that,
for $^{80-82}$Cu, a substantial part to the total $\beta^{-}$-decay
rates came from U1F strength, in line with the conclusion of Borzov
~\cite{Bor05}. The strength of U1F to total $\beta^{-}$-decay rates
decreases, when stellar density increases. However the lepton
capture rates contribution (both allowed GT and U1F) were not
calculated in \cite{Maj17}. The PC rates may compete with the
electron emission rates under stellar conditions. Further the
relative contribution of capture rates to the total stellar weak
rates was also missing in \cite{Maj17}. In the later portion of this
study we compute EC and PC rates for neutron-rich copper isotopes
(71 $\leq$ A $\leq$ 82) employing the
deformed pn-QRPA model in stellar environment.\\
In the next section we  discuss the necessary formalism for
computation of ground-state properties of Cu isotopes using the RMF
model. Here we also briefly explain the necessary pn-QRPA formalism
for the determination of GT strength functions and lepton capture
rates (both allowed GT and U1F). Sect. 3 is devoted to presentation
of our calculated results and associated discussions. Finally, Sect.
4 summarizes our findings.

\section{Theoretical Formalism}

\subsection{The RMF Model}
In the RMF model, a nucleus consists of nucleons and these nucleons
interact with each other in such a way that various mesons and
photons are exchanged between nucleons~\cite{Walecka74}. Scalar
$\sigma$ meson, vector $\omega$ meson and isovector $\rho$ meson are
conventionally taken into account in the RMF model. The $\sigma$
meson is responsible for the attractive part of the interaction of
nucleons while $\omega$ meson is related with the repulsive part.
The photon and $\rho$ meson play key roles for correct description
of electromagnetic interaction and isospin-dependent effects in
nuclei, respectively. Initially interactions of mesons among
themselves were not considered but the simplest version of RMF model
did not account for a correct description of incompressibility for
nuclear matter. For this reason Boguta and Bodmer \cite{Boguta77}
proposed to include a non-linear self interaction of the $\sigma$
mesons in the RMF model. This version of RMF model is commonly known
as the non-linear RMF model and has been used for the last thirty
years for prediction of various nuclear properties of finite nuclei.
Different types of RMF models may be found in literature. In these
models, non-linear self interaction of the $\omega$ and $\rho$
mesons \cite{Sugahara94,Piekarewicz02} as well as density dependent
meson-nucleon couplings \cite{Lenske95,Shen97,lalazissis05} were
considered. In the present study, an effective density-dependent
interaction DD-ME2~\cite{lalazissis05} and a point-coupling
interaction DD-PC1~\cite{niksic08} were considered. For details of
point coupling interaction, the authors suggest to study
Ref.~(\cite{niksic08}). Here we only describe the theoretical
formalism of basic features of the RMF Lagrangian density (given
below in Eq.~(1) with medium dependent vertices).

\begin{eqnarray}
\footnotesize
\mathcal{L}=\bar\psi(i\gamma~.~\partial - m)\psi +
\frac{1}{2} (\partial\sigma)^{2}
-\frac{1}{2}m_{\sigma}^{2}\sigma^{2} -
\frac{1}{4}\boldsymbol{\Omega}_{\mu\nu}\boldsymbol{\Omega}^{\mu\nu}\nonumber\\
+\frac{1}{2}m_{\omega}^{2}\omega^{2} -
\frac{1}{4}\overrightarrow{\boldsymbol{R}}_{\mu\nu}\overrightarrow{\boldsymbol{R}}^{\mu\nu}
+ \frac{1}{2}m_{\rho}^{2}\overrightarrow{\rho}^{2} -
\frac{1}{4}\boldsymbol{F}_{\mu\nu}\boldsymbol{F}^{\mu\nu} - ~g_{\sigma}\bar{\psi}\sigma\psi\\
- g_{\omega}\bar{\psi}\gamma~.~\omega\psi -
g_{\sigma}\bar{\psi}\gamma~.~\overrightarrow{\rho}\overrightarrow{\tau}\psi
- A\frac{1-\tau_{3}}{2}\psi.\nonumber \label{Lagrangian}
\end{eqnarray}

In Eq.~(1), the masses of mesons (fields) are represented by
$m_{\sigma}$ ($\sigma$), $m_{\omega}$ ($\omega$) and $m_{\rho}$
($\rho$). $g_\sigma$, $g_\omega$ and $g_\rho$ are the related
meson-nucleon couplings of these mesons. They are assumed to be
functions of baryon density in practical applications of the
density-dependent hadron field theory. $m$ denotes mass of the
nucleon represented by the Dirac spinor ($\psi$). Bold type symbols
indicate space vectors. Isospin vectors are indicated by arrows. The
Lagrangian given in Eq.~(1) is invariant under parity
transformation. The expectation value of the pseudoscalar pion field
vanishes in  in the mean field approximation because only solutions
with well-defined parity were considered in the study. To reproduce
ground-state properties of nuclei and nuclear matter properties, the
unknown meson masses and coupling constants were adjusted using a
small number of experimental data of few (finite) nuclei. Field
tensors for vector fields $\omega$, $\rho$ and photon in Eq.~(1) are
given by

\begin{eqnarray}
\boldsymbol{\Omega}^{\mu\nu}=\partial^{\mu}\omega^{\nu}-
\partial^{\nu}\omega^{\mu}, \nonumber \label{Omega}\\
\overrightarrow{\boldsymbol{R}}^{\mu\nu}=\partial^{\mu}\overrightarrow{\rho}^{\nu}-
\partial^{\nu}\overrightarrow{\rho}^{\mu}, \label{Rho}\\
\boldsymbol{F}^{\mu\nu}=\partial^{\mu}A^{\nu}-\partial^{\nu}A^{\mu}
\nonumber. \label{Photon}
\end{eqnarray}

By using the Lagrangian density in the classical variational
principle, the equations of motion can be obtained for the fields.
These are a set of coupled equations including Dirac equation for
the nucleons and the Klein-Gordon like equations for mesons and
photons. The resulting equations can be solved for deformed axially
symmetric case \cite{niksic14}. The PECs of $^{71-82}$Cu isotopes
have been obtained from quadrupole moment constrained calculation.
In this calculation the $BE$ at a fixed deformation was computed by
constraining the quadrupole moment $\langle Q_{2} \rangle$ to a
given value $\mu_{2}$ in the expectation value of the Hamiltonian
given as
\begin{equation}
\langle H^{\prime} \rangle = \langle H \rangle + C_{\mu} (\langle
Q_{2} \rangle - \mu_{2})^{2}. \label{Constraining}
\end{equation}
$C_{\mu}$ is the constraint multiplier in this equation. The
relation between the deformation parameter $\beta_{2}$ and the
expectation value of quadrupole moment $\langle Q_{2} \rangle$ was
taken as
\begin{equation}
\langle Q_{2} \rangle=(3/\sqrt{5\pi})Ar^{2}\beta_{2}
\label{Quadrupole}
\end{equation}
where $r=R_{0}A^{1/3}$ ($R_{0}=1.2$ fm) and $A$ is the mass number
of the nucleon. For pairing correlations, Bardeen-Cooper-Schrieffer
(BCS) method was considered and the constant \textit{G}
approximation was used. As proposed by Karatzikos et al.
\cite{Karatzikos10}, a fixed pairing strength $G$ was kept in our
calculation.

\subsection{The pn-QRPA Model}
We calculate  the super allowed Fermi, allowed GT $\&$ U1F
transitions and stellar lepton capture rates using the pn-QRPA
model. The Coulomb interaction with the nucleus distorts the
electron wave functions and are characterized by the corresponding
Fermi functions in the phase space integrals. We assume that the
stellar temperature is sufficiently high due to which the electrons
are not be bound to the nucleus. They are described by the
Fermi-Dirac energy distribution function. At high stellar
temperatures ($kT
> $ 1 MeV), positrons are created via electron-positron pair
production. We  assume that the positrons follow the same energy
distribution as the electrons. We later consider the capturing of
these positrons in our calculation (a process that competes with the
$\beta$-decay rates presented earlier in \cite{Maj17}). We further
assume in our calculation that, for the range of stellar density
considered in this work, the neutrinos and antineutrinos (produced
due to lepton capture reactions) escape freely from the stellar
 core (neutrino capture is not considered in our calculation).

The deformed pn-QRPA Hamiltonian may be written as

\begin{equation} \label{GrindEQ__1_}
H^{pn-QRPA} =H^{Nilsson} +V^{BCS} +V_{GT}^{ph} +V_{GT}^{pp},
\end{equation}
where $H^{Nilsson}$ is the single-particle Hamiltonian in a deformed
Nilsson basis, $V^{BCS}$ represents the pairing interaction solved
using the BCS approach and the last two terms $V_{GT}^{ph}$ and
$V_{GT}^{pp}$  are the particle-hole ($ph$) and particle particle
($pp$) GT forces. Nilsson approach \cite{Nil55} was utilized for the
calculation of wave functions and single particle energies taking
into consideration the nuclear deformation. BCS approximation was
carried out for the protons and neutrons, separately, in the
deformed Nilsson basis. In our model, the residual interaction
occurs in $ph$ and $pp$ channels. The interactions were given
separable form and their respective strength were controlled by
corresponding model parameters ($\chi$ for $ph$ force and $\kappa$
for $pp$ force). The model parameters were selected in order to best
reproduce the experimental half-lives. Our calculation obeyed the
model independent (ISR) sum rule \cite{isr63}.  The value of $\chi$,
taken for both allowed GT and U1F, was 61.20/A. However for allowed
GT, $\kappa =$ 4.85/A (MeV), and for the U1F case, $\kappa =$
10.92/A (MeV fm$^{-2}$) were considered in our calculation. The
selected values of $\chi$ and $\kappa$ exhibited an inverse mass
dependence as suggested in Refs. \cite{Hom96,Nab15,Nab16,Nab17}.
Other variables for the capture rates estimation are the pairing gap
($\Delta _{p}$, $\Delta _{n}$), nuclear deformation ($\beta_{2}$),
Q-values and the Nilsson potential parameters (NPP). The $\beta_{2}$
values were taken from Ref.~\cite{Mol95}. Earlier the stellar
electron emission rates for the selected Cu isotopes were computed
\cite{Maj17} using deformation from Ref.~\cite{Mol95}. We used
deformations from same reference in order to compare our calculated
positron capture with the previous calculated $\beta$-decay rates in
stellar matter which we discuss later. The NPP were adopted from
~\cite{rag84} and the oscillation constant (identical for both
protons and neutrons) was determined using the relation $\hbar
\omega=41A^{-1/3}$ (MeV). Pairing gap value of $\Delta_n = \Delta_p
= 12/\sqrt A$ (MeV) was employed in our calculation. For the
determination of Q-values we used the mass compilation of Audi and
collaborators \cite{aud12}.

The weak lepton capture rates were calculated for the following two
charge-changing transitions on $^{71-82}$Cu:\\
1)\hspace{0.2in}Electron Capture:\hspace{0.5in}
$^{A}_{29}$Cu+e$^{-}$ $\longrightarrow$ $^{A}_{28}$Ni+$\nu_{e}$
\newline
\\
2)\hspace{0.2in}Positron Capture:\hspace{0.5in}
$^{A}_{29}$Cu+e$^{+}$ $\longrightarrow$
$^{A}_{30}$Zn+$\bar{\nu_{e}}$
\\
\newline
The allowed EC/PC weak-rates from the parent $k^{th}$ level to the
daughter $l^{th}$ level of the nuclide are given by
\begin{eqnarray}\label{lkl}
\lambda_{EC(PC)} ^{kl} =\left[\frac{\ln 2}{D}
\right]\left[f^{kl} (T,\rho ,E_{f} )\right] \nonumber \\
\left[B(F)^{kl} +\left({\raise0.7ex\hbox{$ g_{A} $}\!\mathord{\left/
{\vphantom {g_{A}  g_{V} }} \right.
\kern-\nulldelimiterspace}\!\lower0.7ex\hbox{$ g_{V}  $}}
\right)^{2}_{eff} B(GT)^{kl} \right].
\end{eqnarray}
We took value of D to be 6143s \cite{Har09}. In Eq.~(\ref{lkl})
$B^{kl}$'s represent sum of reduced transition probabilities of the
Fermi B(F) and GT transitions B(GT). The value of $(g_{A}/g_{V})$,
denoting the ratio of axial and vector coupling constants, was taken
as -1.2694 \cite{Nak10}. The detailed formalism for calculation of
capture rates for allowed transitions under stellar scenario may be
seen from Ref. \cite{nab99}.

The U1F stellar lepton capture rates from the parent $k^{th}$ level
to the daughter nuclide $l^{th}$ level are given by

\begin{equation}\label{lij}
\lambda^{kl}_{EC(PC)} =
\frac{m_{e}^{5}c^{4}}{2\pi^{3}\hbar^{7}}\sum_{\Delta
J^{\pi}}g^{2}f^{kl}(\Delta J^{\pi})B^{kl}(\Delta J^{\pi}),
\end{equation}
where $f^{kl}(\Delta J^{\pi})$ and $ B^{kl}(\Delta J^{\pi})$ are the
Fermi function and the reduced transition probability for the
capture processes. For necessary formalism of calculation of U1F
transitions in the deformed pn-QRPA model with separable interaction
see Refs.~\cite{Nab16,Nab17,Maj17}.

The large temperature inside the core of massive stars implies that
there is a finite chance of occupations of parent excited levels in
stellar matter. The total EC ($\lambda _{EC}$) and positron capture
($\lambda _{PC}$) rates per unit time per nucleus are finally given
by
\begin{equation}
\lambda _{EC(PC)} =\sum _{kl}P_{k} \lambda _{EC(PC)}^{kl},
\label{lb}
\end{equation}
where $P_{k}$ represents the probability of occupation of parent
excited level and is determined using the normal Boltzmann
distribution. The sum incorporated in Eq.~(\ref{lb}) is taken on all
the initial and final levels for calculation of total capture rates.
Convergence in our capture rate calculations was ensured (we used a
large model space of up to the 7$\hbar\omega$ in our pn-QRPA
calculation).

\section{Results and comparison}

As mentioned earlier, we employed the RMF model, with density
dependent DD-ME2 and DD-PC1 interactions, to calculate the
ground-state nuclear properties of $^{71-82}$Cu nuclide. The
calculated nuclear properties include binding energy per nucleon
($BE/A$), radii of neutron and proton, root-mean-square charge
radius, electric quadrupole moment and quadrupole deformation
parameter ($\beta_{2}$). We also present the ground-state shape
evolution of $^{71-82}$Cu isotopes based on their PECs.

\begin{center}
\begin{figure*}[h]
\centering
\includegraphics[width=0.6\textwidth]{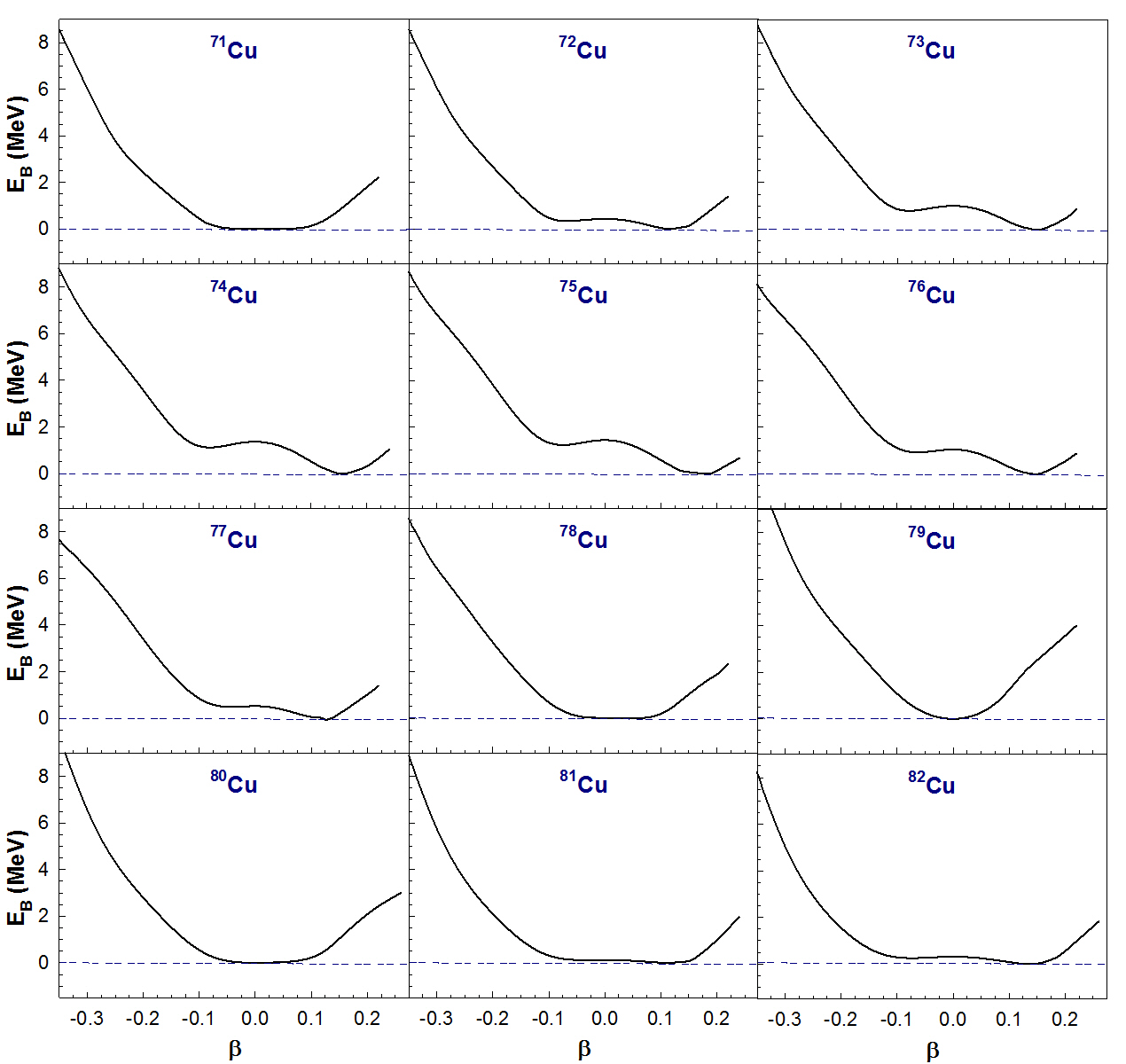}
\caption{(Color online) The potential energy curves for $^{71-82}$Cu
obtained using the quadrupole moment constrained RMF model with the
DD-ME2 interaction.} \label{pec_me2}
\end{figure*}
\end{center}

\begin{figure*}[h]
\centering
\includegraphics[width=0.6\textwidth]{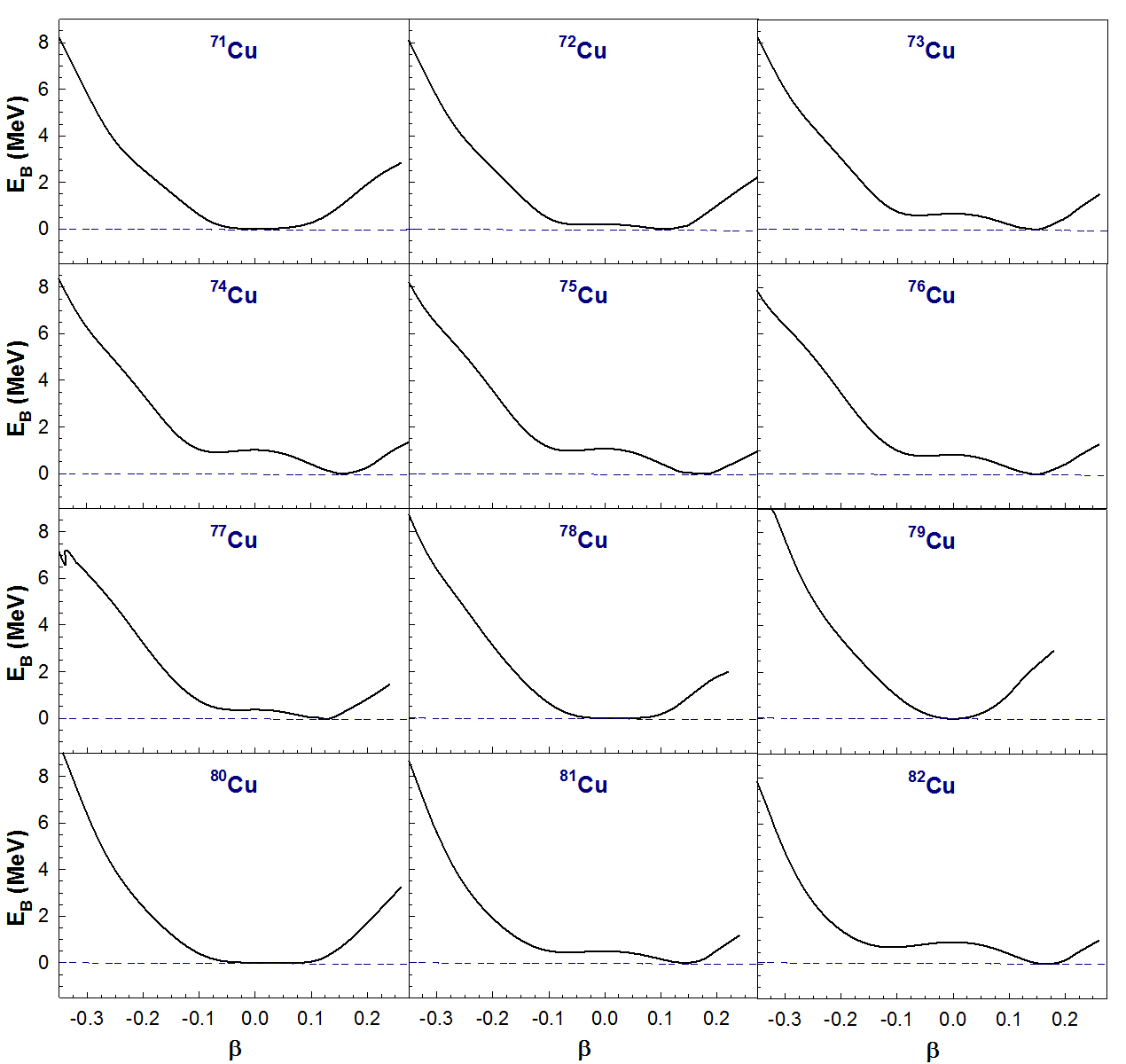}
\caption{(Color online) The potential energy curves for $^{71-82}$Cu
obtained using the quadrupole moment constrained RMF model with the
DD-PC1 interaction.} \label{pec_pc1}
\end{figure*}

\begin{figure*}[h]
\centering
\includegraphics[width=0.6\textwidth]{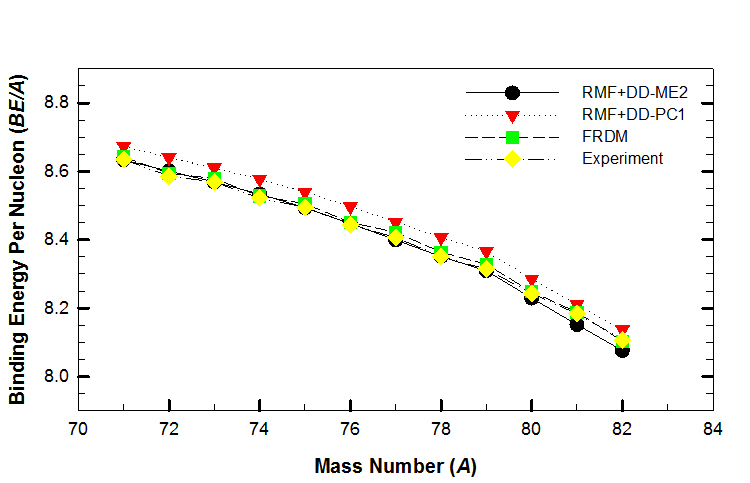}
\caption{(Color online) Binding energy per nucleon for $^{71-82}$Cu.
The predictions of the RMF model, using DD-ME2 and DD-PC1
interactions, are compared with the FRDM calculation
\cite{moller1997} and experimental \cite{aud12} results.}
\label{bea}
\end{figure*}

The calculated PECs from our quadrupole moment constrained RMF
calculation, using DD-ME2 and DD-PC1 functionals for $^{71-82}$Cu,
are displayed in Figures ~\ref{pec_me2} and \ref{pec_pc1},
respectively. The ground state binding energy of each isotope is
taken as reference in these figures. $^{79}$Cu has shell closure
with magic neutron number N = 50 and one can expect that it has
spherical shape. This is clearly visible in the PECs of $^{79}$Cu
both for DD-ME2 and DD-PC1 functionals. It should be noted that
neighboring isotopes of closed shell and semi-closed shell nuclei
may exhibit spherical character in RMF model. The neutron number N =
40 is known as semi-magic number. $^{71}$Cu  has neutron number N =
42 which is close to semi-magic neutron number N = 40. Because of
this reason, the ground-state shape of $^{71}$Cu is seen as only
slightly deformed in both  Fig.~\ref{pec_me2} and
Fig.~\ref{pec_pc1}. The shape of isotopes starting from $^{72}$Cu to
$^{77}$Cu are seen as prolate in these figures. $^{78}$Cu and
$^{80}$Cu close to neutron shell closure have only slightly deformed
shapes. Finally the shapes of ground-state of $^{81}$Cu and
$^{82}$Cu become prolate again. It is noted from Figs.~\ref{pec_me2}
and \ref{pec_pc1} that similar PECs are obtained in RMF model with
DD-ME2 and DD-PC1 interactions. Based on these PECs we obtained
DD-ME2 and DD-PC1 functional predictions for the ground-state $BE/A$
of $^{71-82}$Cu nuclei. The results are shown in Fig.~\ref{bea}.
Here we also show the FRDM predictions \cite{moller1997} and
experimental data \cite{aud12} for the sake of comparison. The
calculated $BE/A$ for $^{71-82}$Cu using the DD-ME2 interaction in
RMF model and the predictions of FRDM model are in agreement with
the experimental data. The predictions of DD-PC1 for the $BE/A$ of
$^{71-82}$Cu are slightly different when they are compared with the
results of DD-ME2 interaction. However, the maximal deviation
between experimental data and the calculated $BE/A$ using the DD-PC1
interaction is about $0.056$ MeV. It should be noted that FRDM gives
good results for ground-state binding energies of nuclei throughout
the nuclidic chart by fitting many parameters. On the other hand the
RMF model, with smaller number of parameters, shows result at par
with the  FRDM model \cite{bayram2013a}. The root-mean-square (rms)
deviation between the calculated values of $BE/A$ using FRDM,
RMF+DD-ME2 and RMF+DD-PC1 models and the experimental $BE/A$ values
are 0.726, 1.195 and 3.490, respectively. It is concluded that RMF
model with DD-ME2 interaction is closer to the FRDM results for
predictions of the ground-state $BE/A$ of $^{71-82}$Cu isotopes.

\begin{figure*}[h]
\centering
\includegraphics[width=0.6\textwidth]{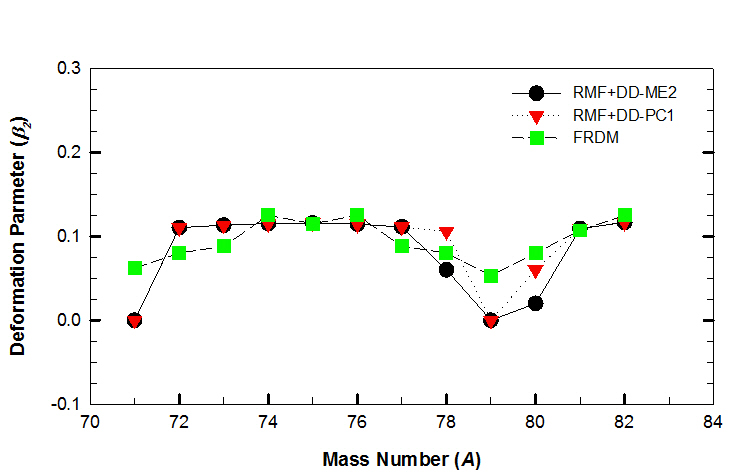}
\caption{(Color online) Calculated quadrupole deformation parameter
($\beta_{2}$) for $^{71-82}$Cu. The predictions of the RMF model,
using DD-ME2 and DD-PC1 interactions, are compared with those of
FRDM \cite{moller1997}.} \label{beta}
\end{figure*}

\begin{figure*}[h]
\centering
\includegraphics[width=0.6\textwidth]{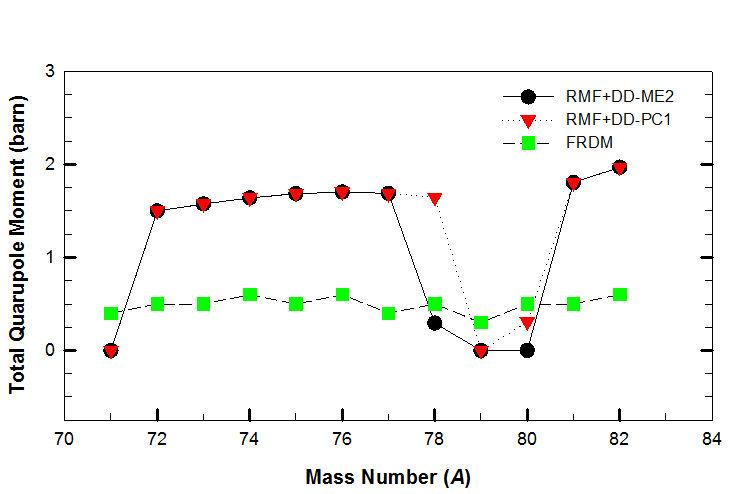}
\caption{(Color online) Total quadrupole moment for $^{71-82}$Cu.
The predictions of the RMF model, using DD-ME2 and DD-PC1
interactions, are compared with the those of FRDM
\cite{moller1997}.} \label{qt}
\end{figure*}

Nuclear deformation is one of the important nuclear properties. The
RMF model can reproduce the deformations of finite nuclei rather
well \cite{lalazissis1996,lalazissis1999}. In Fig.~\ref{beta}, the
calculated quadrupole deformation parameters ($\beta_{2}$) for
$^{71-82}$Cu using the DD-ME2 and DD-PC1 interactions are shown in
comparison with those calculated using the FRDM. As can be seen from
Fig.~\ref{beta}, the calculated values of $\beta_{2}$ using the RMF
model with DD-ME2 and DD-PC1 functionals for Cu isotopes are close
to zero around neutron number $N=50$ which implies that RMF model
predicts the shape of nuclei as spherical at N = 50. Also, a similar
situation can be seen in Fig.~\ref{qt}. In this figure, the
calculated values of total (neutron+proton) electric quadrupole
moments ($Q_{T}$) are shown for $^{71-82}$Cu isotopes. The results
of DD-ME2 and DD-PC1 functionals are close to zero at and near the
neutron number $N$= 50. It is observed that the computed values of
electric quadrupole moments obtained from RMF model is higher than
those from FRDM.

We calculated neutron and proton radii ($r_{n}$ and $r_{p}$) of
$^{71-82}$Cu isotopes in our study by considering the DD-ME2 and
DD-PC1 interactions. Similar results have been obtained in our
calculations for these interactions. Root mean square (rms) charge
radius ($r_{c}$) of nuclei can be calculated by inserting proton
radius ($r_{p}$) in the formula $r_{c} = \sqrt{r_{p}^{2} + 0.64}$
fm. To the best of our knowledge there is no available experimental
charge radii data for $^{71-82}$Cu. We have listed the calculated
rms charge radii, neutron and proton radii of $^{71-82}$Cu, using
the DD-ME2 interaction, in Table~\ref{Table 1}. The results of our
RMF model calculation for $BE/A$, $\beta_{2}$ and $Q_{T}$ are also
listed in this table.

In the remaining portion of this section we present discussion on
the results obtained from the pn-QRPA model. These results are
important for astrophysical applications. We used a quenching factor
(f$_{q}$) of 0.6 in present pn-QRPA calculation (the same $f_{q}$
value was suggested for the RPA results in case of $^{54}$Fe nuclide
\cite{vet89}).

\begin{figure*}[h]
\centering
\includegraphics[width=0.7\textwidth]{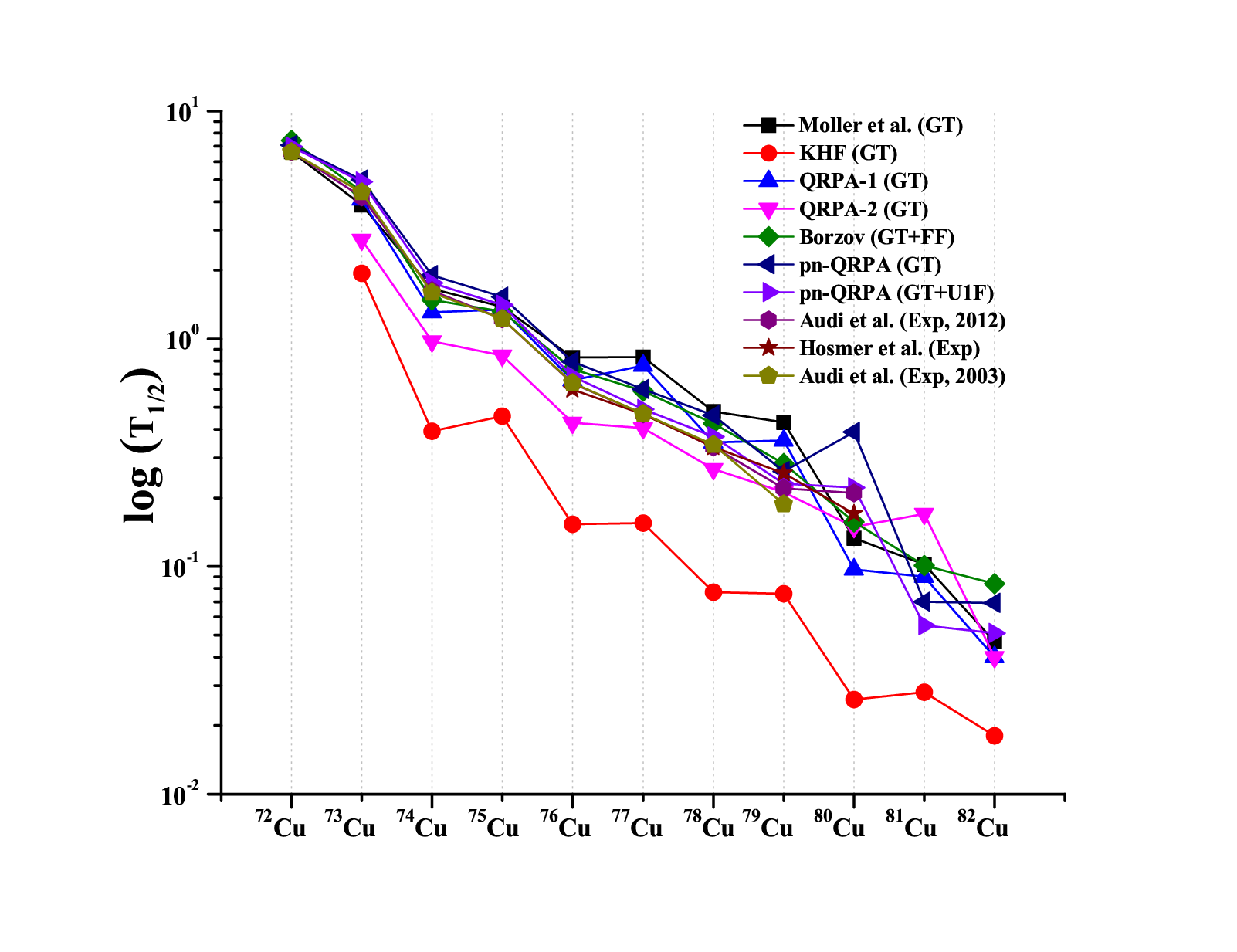}
\caption{(Color online) Calculated terrestrial half-lives
($T_{1/2}$) for copper isotopes using the pn-QRPA model (this work)
in comparison with previous theoretical calculations and
experimental results. For references see text.} \label{fig6}
\end{figure*}

To check the reliability of the current pn-QRPA model, we first
discuss and compare our calculated terrestrial half-lives
($T_{1/2}$) of copper isotopes with previous measured data and
theoretical calculations (Fig.~\ref{fig6}). The experimental
half-lives were taken from \cite{aud03,aud12} and
 ~\cite{Hos10}. Recently, authors in \cite{Hos10}
measured the T$_{1/2}$ values for neutron-rich nuclei and it was
concluded that the nuclear deformation parameter ($\beta_2$) have a
sizable contribution to the T$_{1/2}$ values. The authors in
\cite{moller1997} used the FRDM + RPA model, by considering only
allowed transitions (indicated as M\"{o}ller et al. in
Fig.~\ref{fig6}). Borzov \cite{bor05} used the (DF3 + CQRPA) model
for the calculation of allowed and U1F charge-changing transitions,
by considering spherical nuclide. It was found by Hosmer et al. that
the spherical shape assumption considered in Borzov work is not
justified. KHF, QRPA-1 and QRPA-2 are the results of \cite{pfe02}
calculation, in which only allowed GT rates were computed. The
deformed pn-QRPA computed half-lives, with and without U1F
contribution, are also displayed in Fig.~\ref{fig6}. The overall
comparison shows that the deformed pn-QRPA model (this work)
reproduces well the measured $T_{1/2}$ values as compared to other
theoretical models. It is seen that the addition of U1F rates to the
allowed GT rates further improve the comparison of our computed
half-lives with the measured results. It is noted that the
comparison of calculated T$_{1/2}$ values may be improved further by
integrating rank 0 and 1 operators (non-unique in nature) in our
calculation which we plan to investigate in the near future.

\begin{figure*}[h]
\begin{center}
\begin{tabular}{cc}
\centering
\includegraphics[width=0.47\textwidth]{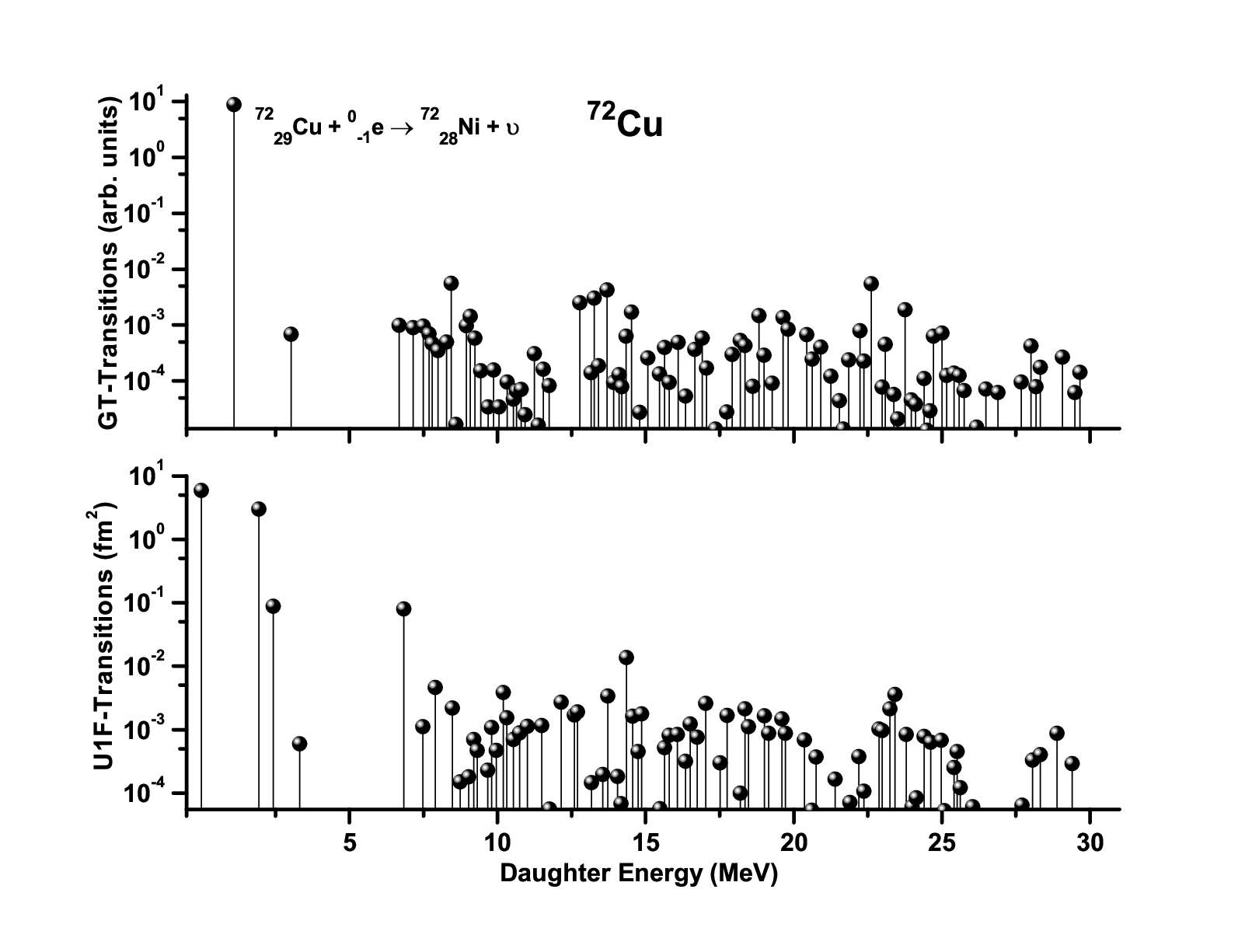}&
\includegraphics[width=0.47\textwidth]{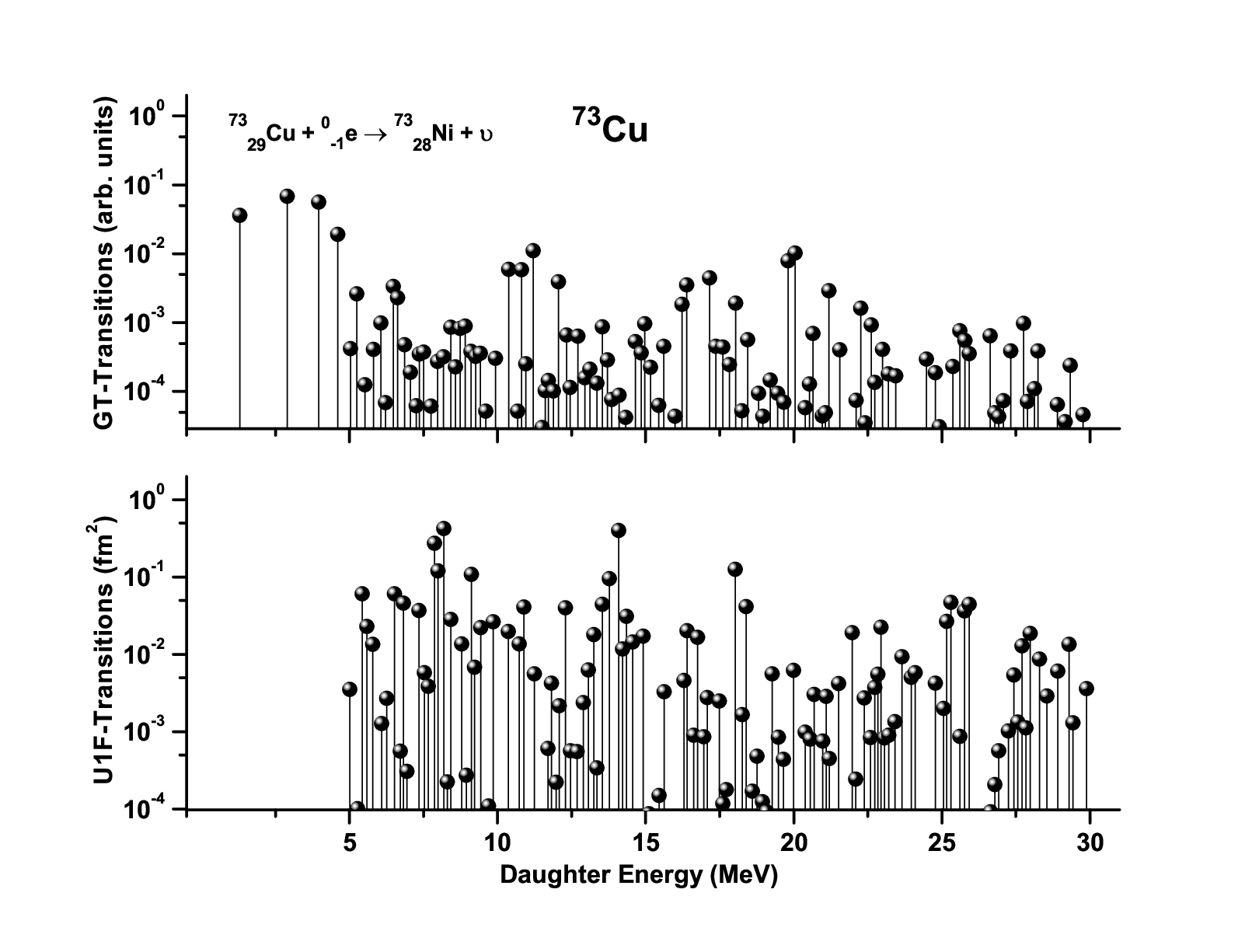}\\
\includegraphics[width=0.47\textwidth]{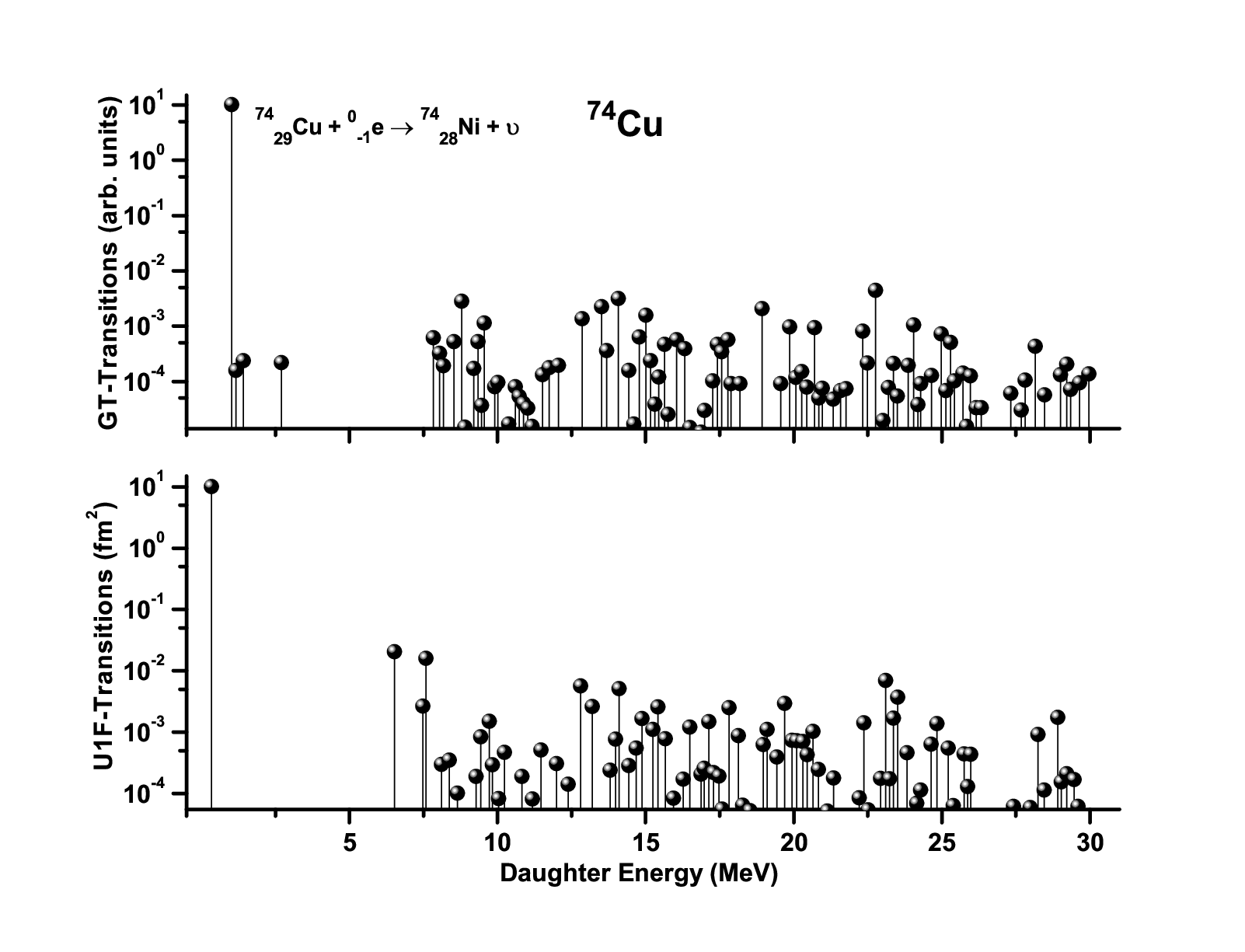}&
\includegraphics[width=0.47\textwidth]{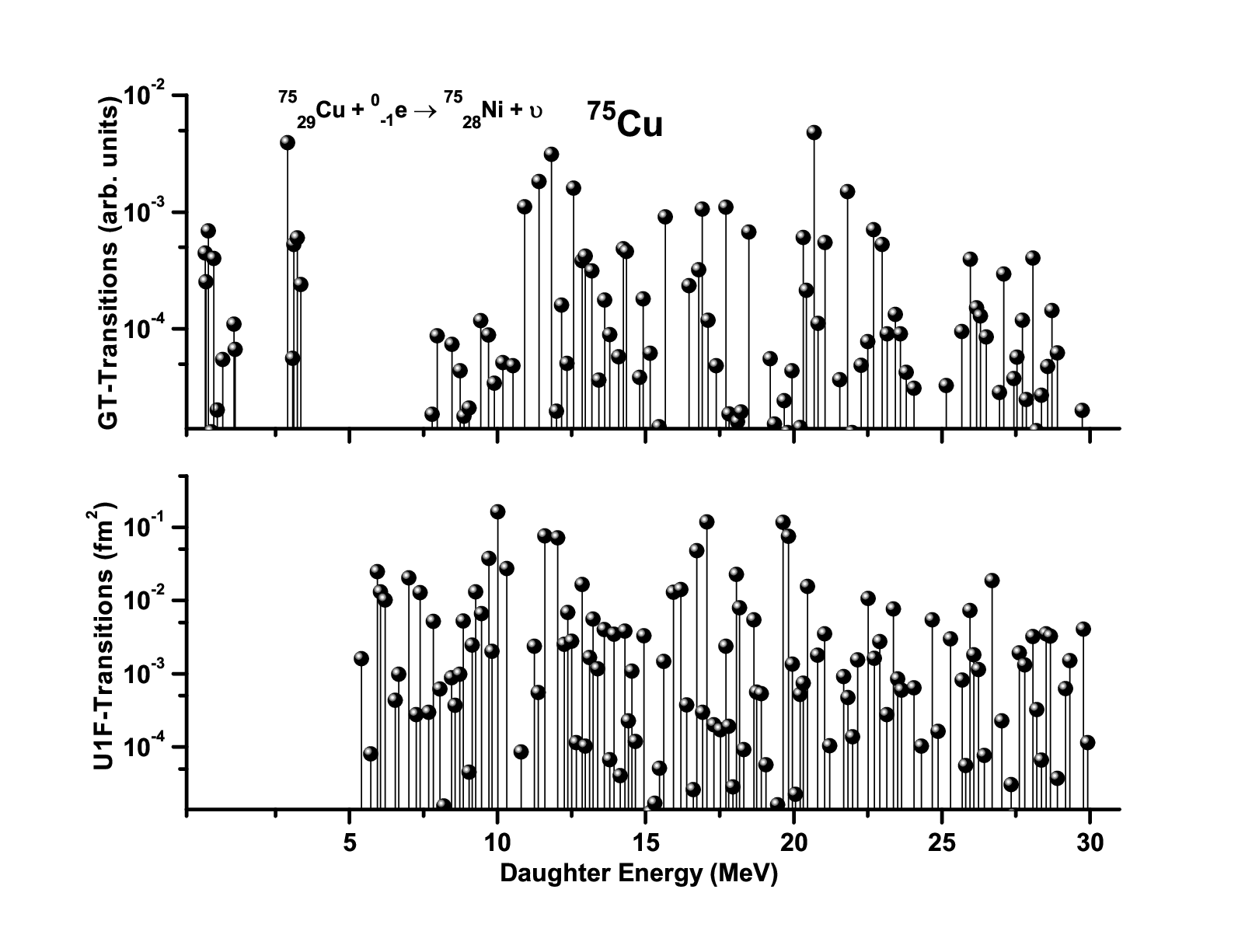}\\

\end{tabular}
\caption{The pn-QRPA calculated allowed and U1F transitions for
$^{72-75}$Cu  as a function of daughter excitation energy in
electron capture direction.} \label{figure7}
\end{center}
\end{figure*}

\begin{figure*}[h]
\begin{center}
\begin{tabular}{cc}
\centering
    \includegraphics[scale=0.31]{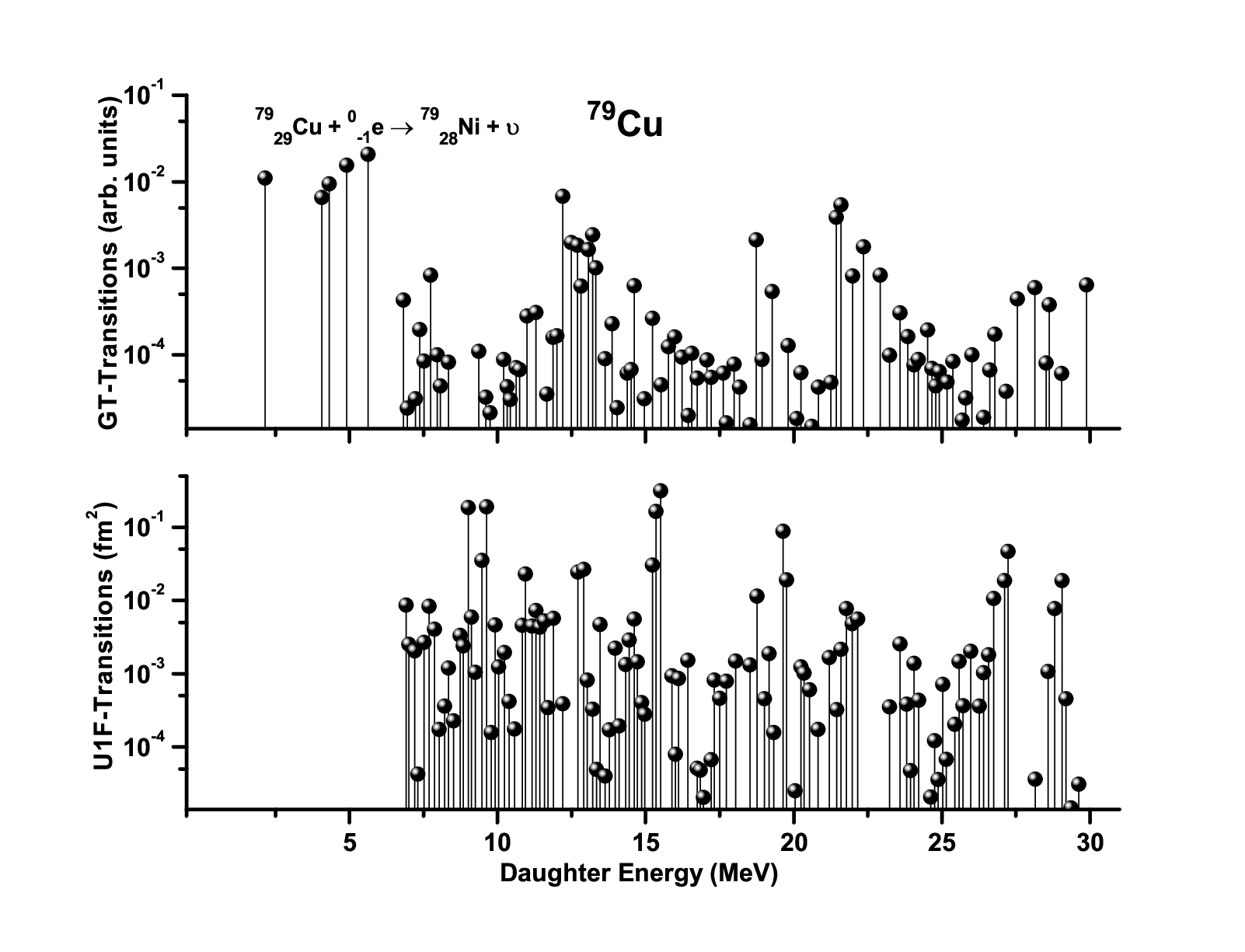} &
    \includegraphics[scale=0.31]{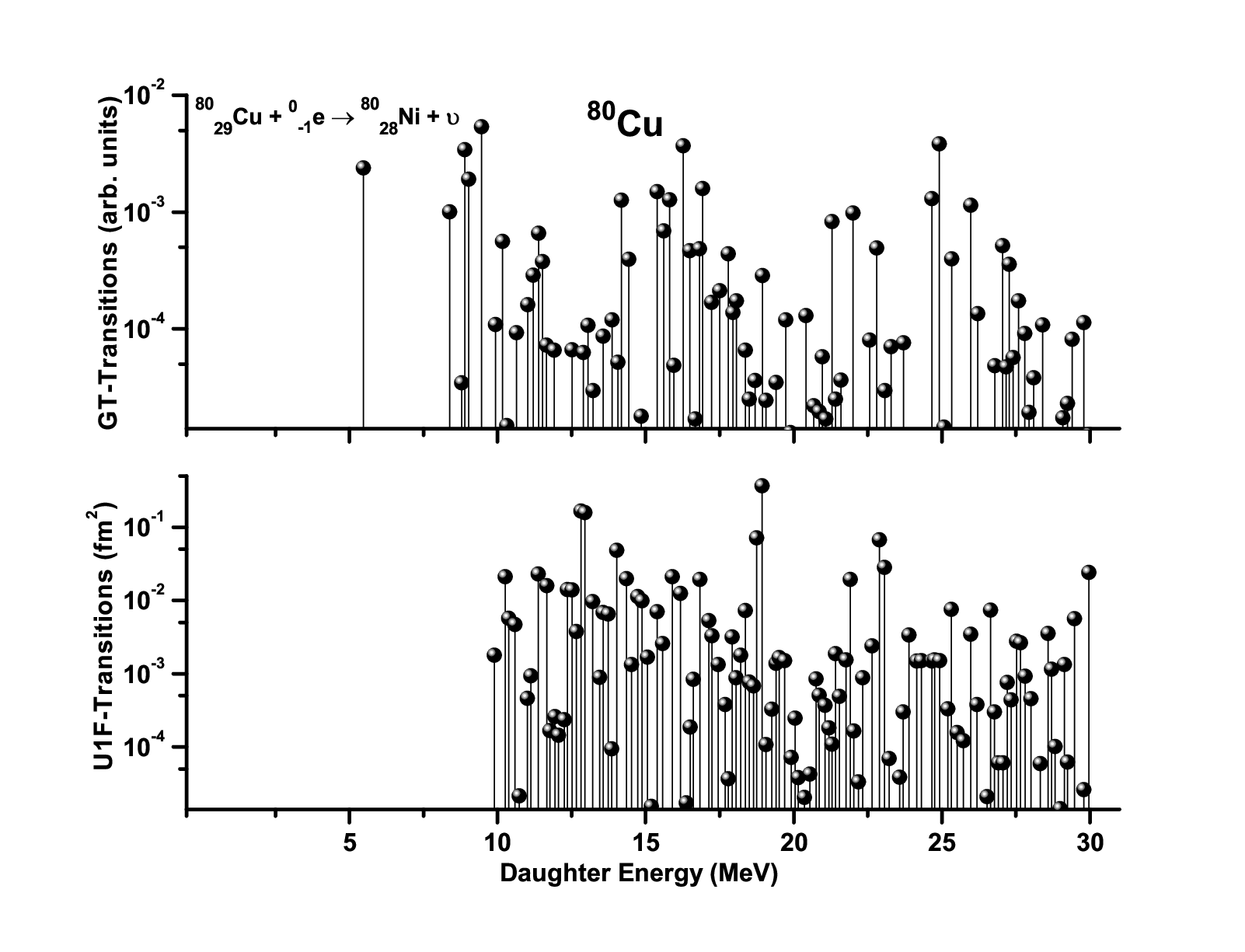}\\
    \includegraphics[scale=0.31]{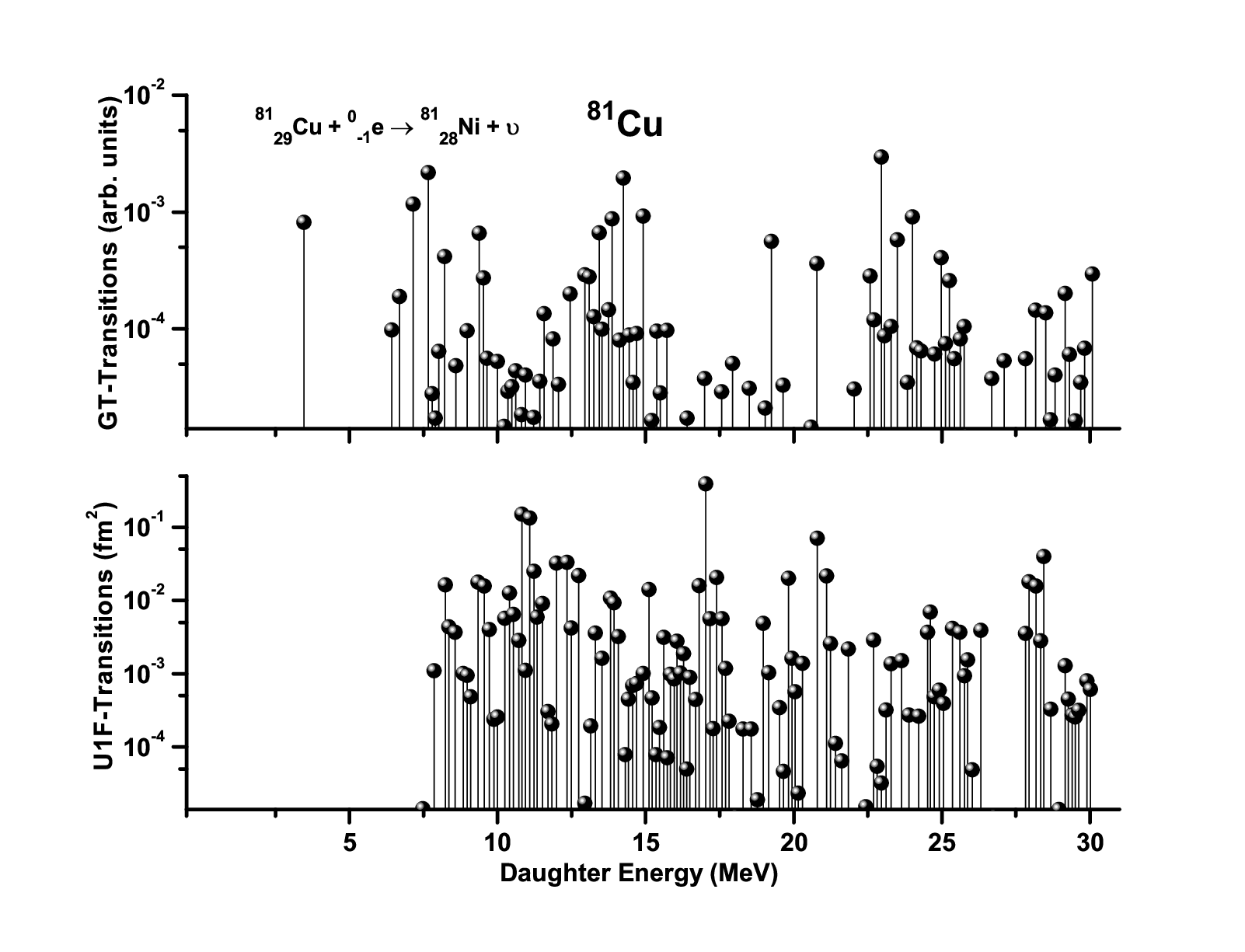} &
    \includegraphics[scale=0.31]{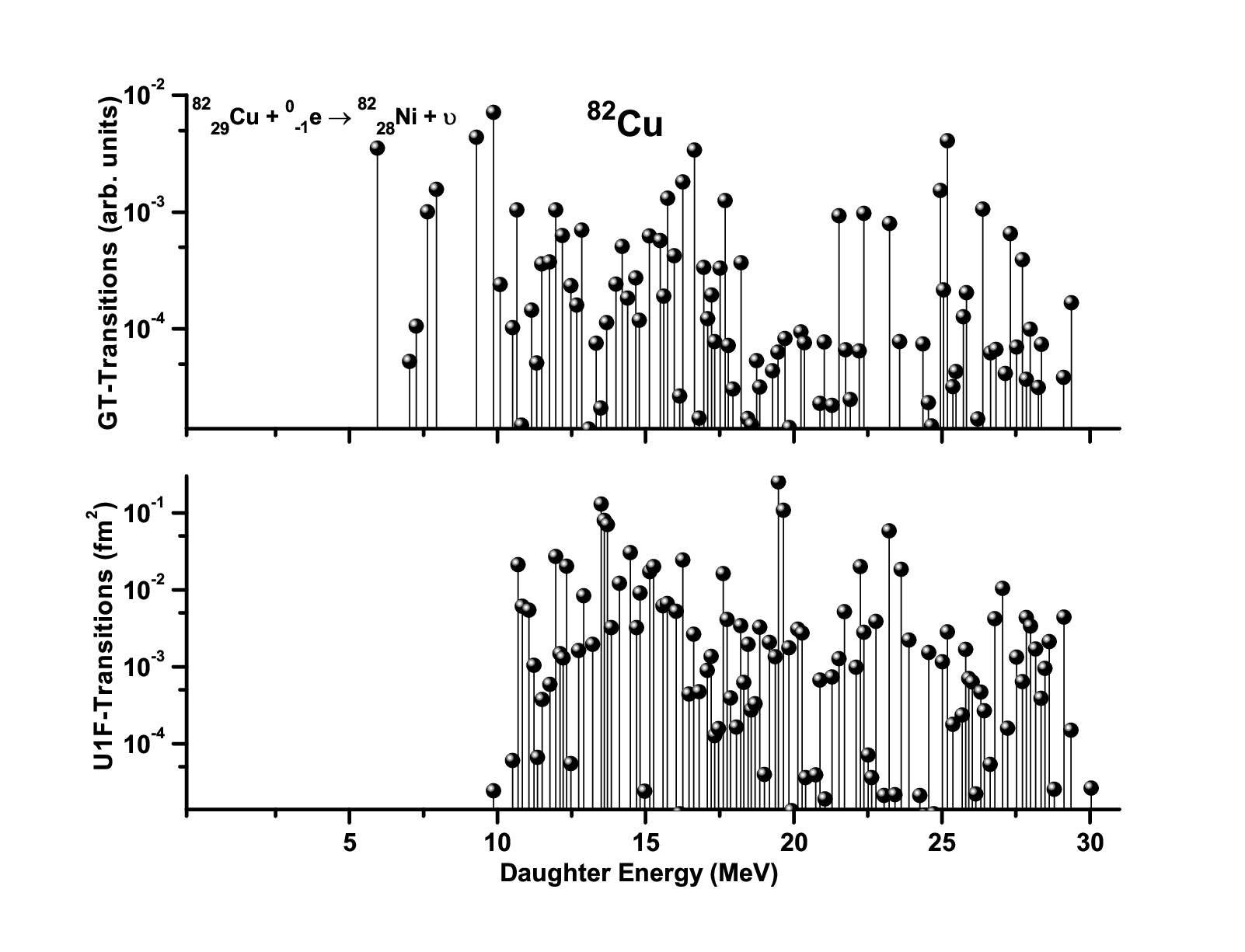}\\
\end{tabular}
\caption{Same as Fig.~\ref{figure7}, but for $^{79-82}$Cu.}
\label{figure8}

\end{center}
\end{figure*}

We present the computed allowed GT and U1F strength distributions
for neutron-rich Cu nuclei in Fig.~\ref{figure7} and
Fig.~\ref{figure8}. The allowed GT strength are shown in arbitrary
units while U1F strength are displayed in fm$^{2}$ units.  The
strength distributions are plotted up to excitation energy of 30 MeV
in daughter nuclide. Charge-changing strength of magnitude less than
10$^{-5}$, though calculated, are not shown in Fig.~\ref{figure7}
and Fig.~\ref{figure8}. To further augment the reliability of our
calculated lepton capture rates, we integrated the experimental
energy levels (XUNDL) in our computation. The pn-QRPA computed
excitation energy levels were swapped with the experimental levels
when they were within 0.5 MeV of one another. Missing measured
energy levels were augmented together with their $\log$\emph{ft}
values wherever appropriate. Computed levels remain unchanged beyond
experimental states for which spin and/or parity assignments were
uncertain. Nuclear deformation was taken into account in the current
pn-QRPA model which resulted in the fragmentation of GT and U1F
transition strengths as shown in the figures. For the heavy copper
isotopes it is noted that U1F transitions appear well above 5 MeV in
daughter level (Fig.~\ref{figure8}). Excited state GT and U1F transitions
were also calculated but not presented due to space limitations. The
ASCII files of all allowed GT and U1F strength distributions may be
requested from the corresponding author.

\begin{figure*}[h]
\begin{center}
\begin{tabular}{cc}
\centering
\includegraphics[width=0.47\textwidth]{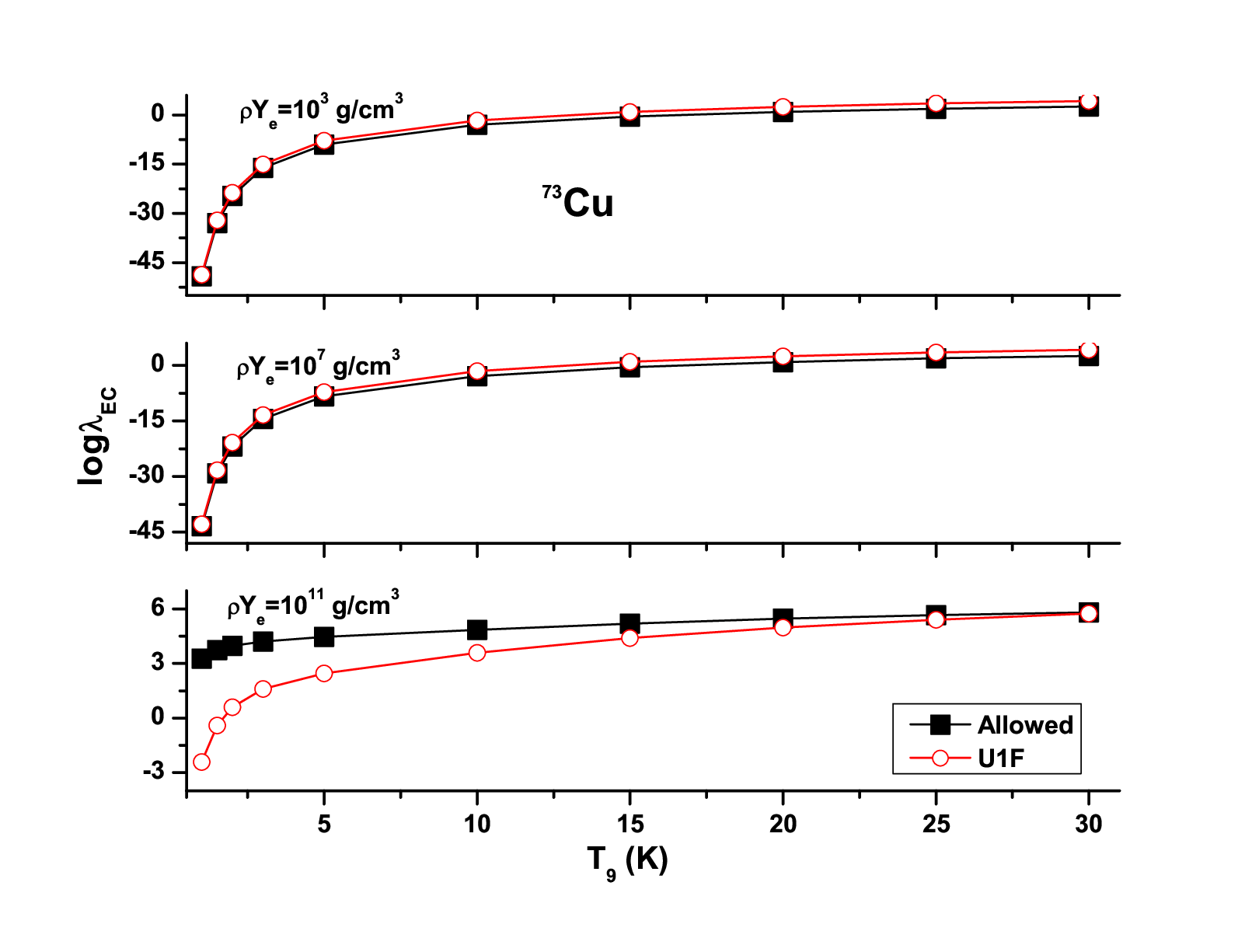}&
\includegraphics[width=0.47\textwidth]{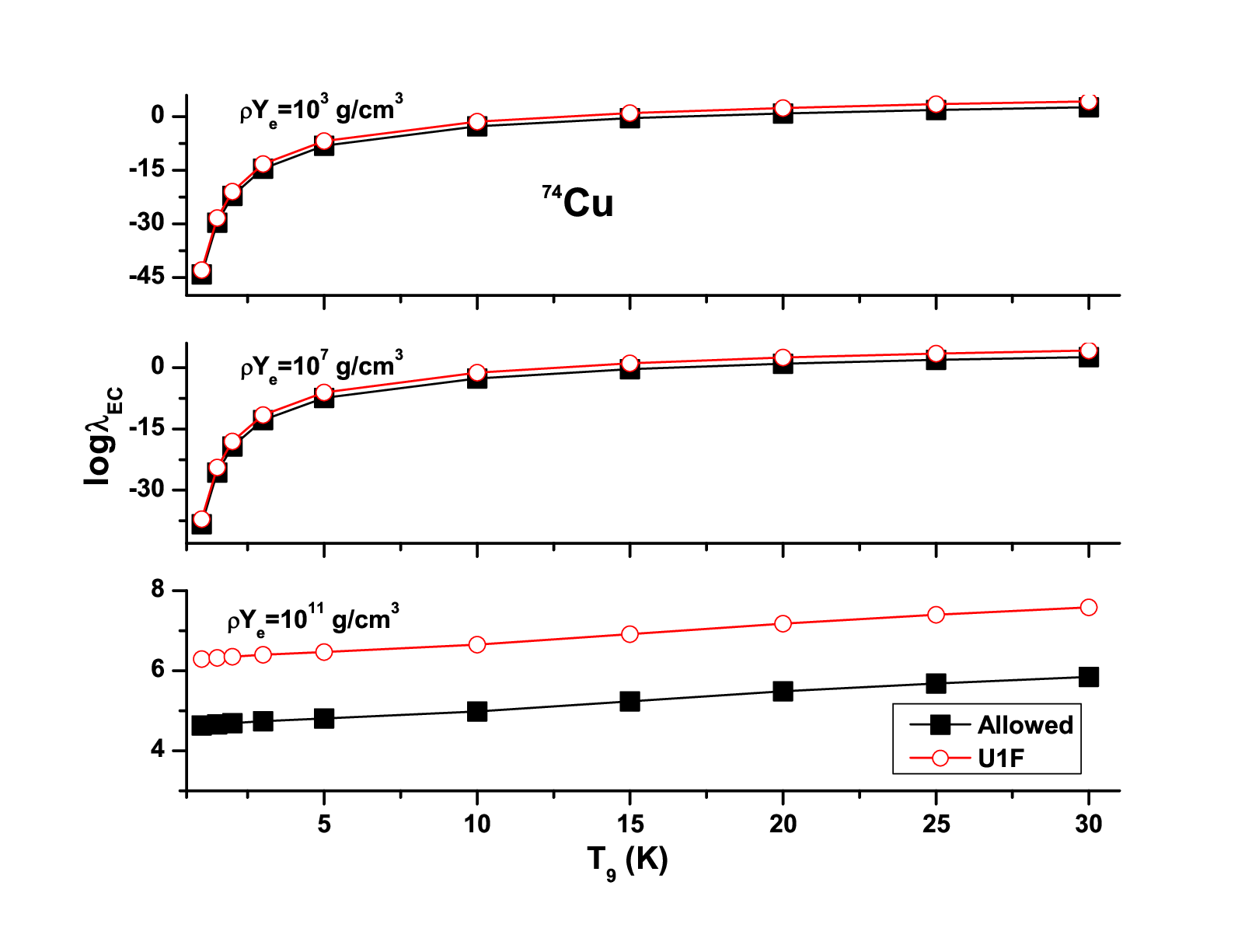}\\
\includegraphics[width=0.47\textwidth]{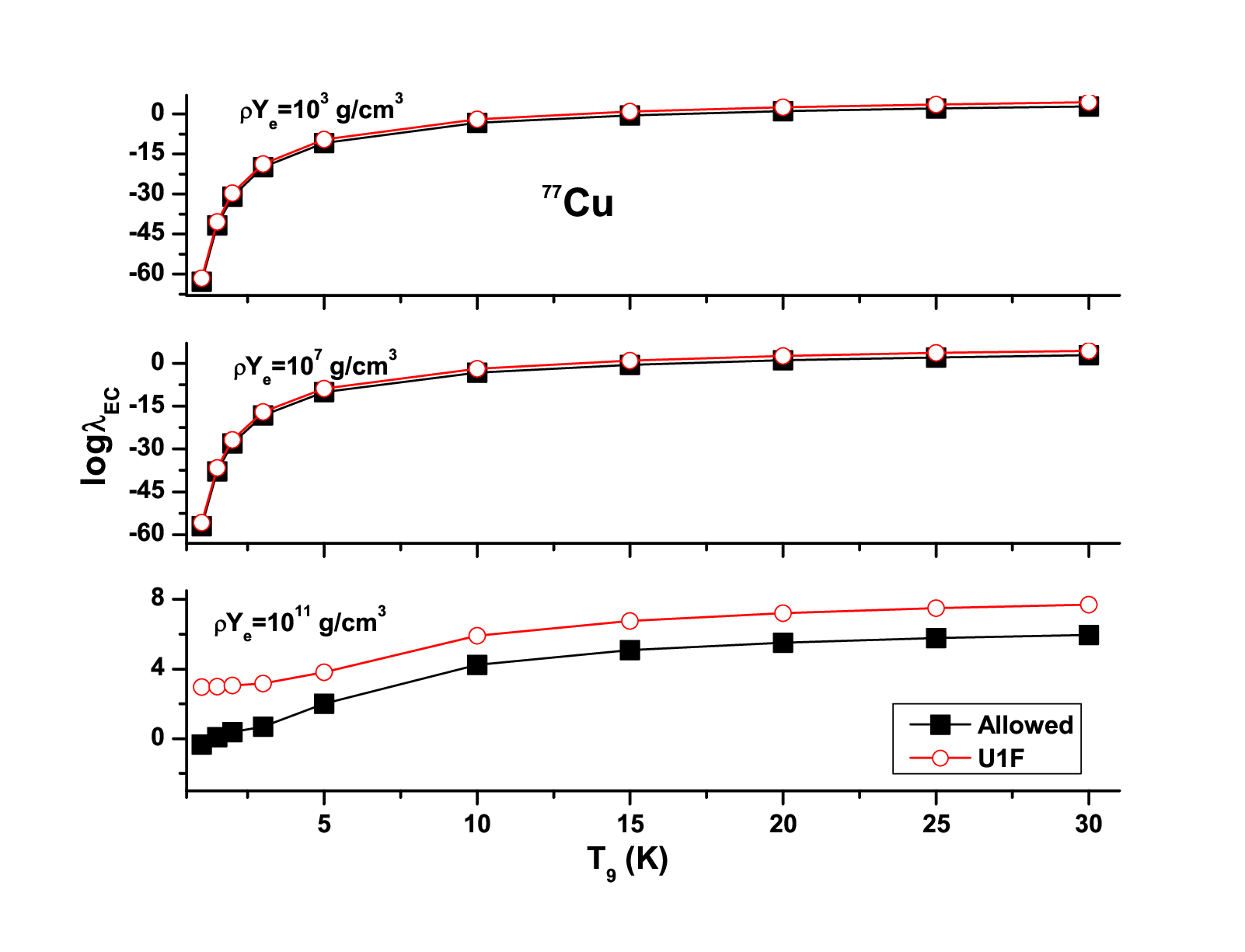}&
\includegraphics[width=0.47\textwidth]{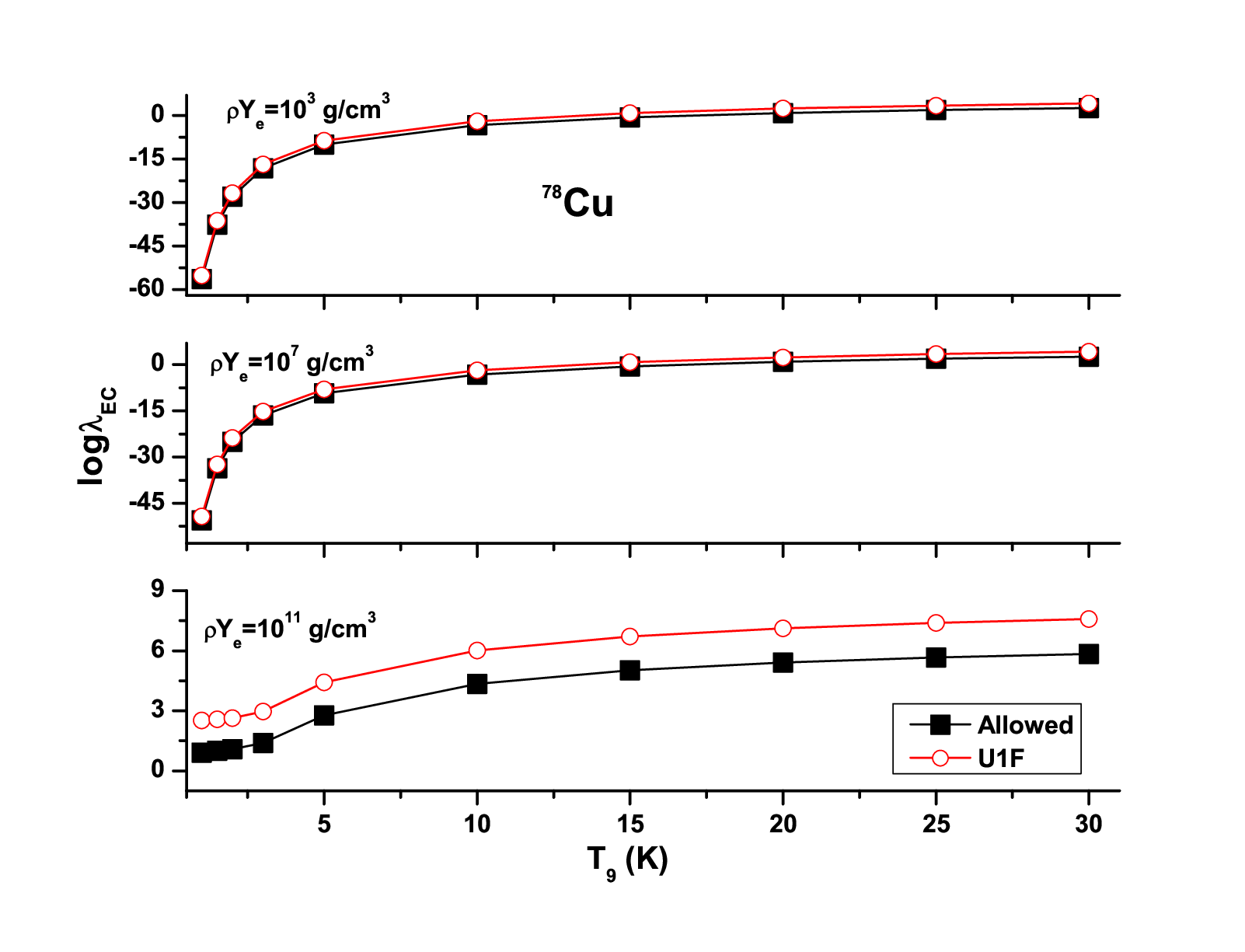}\\
\includegraphics[width=0.47\textwidth]{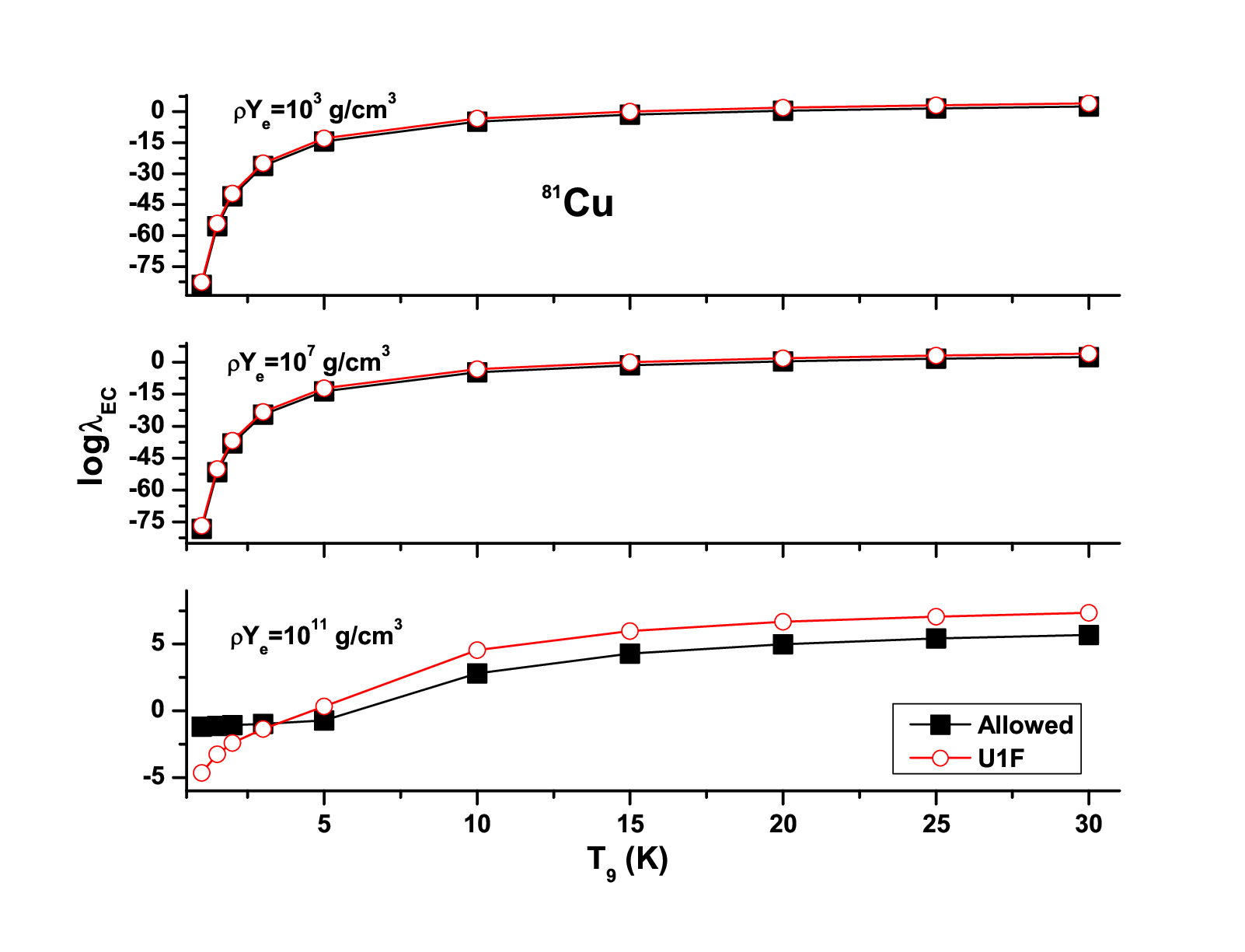}&
\includegraphics[width=0.47\textwidth]{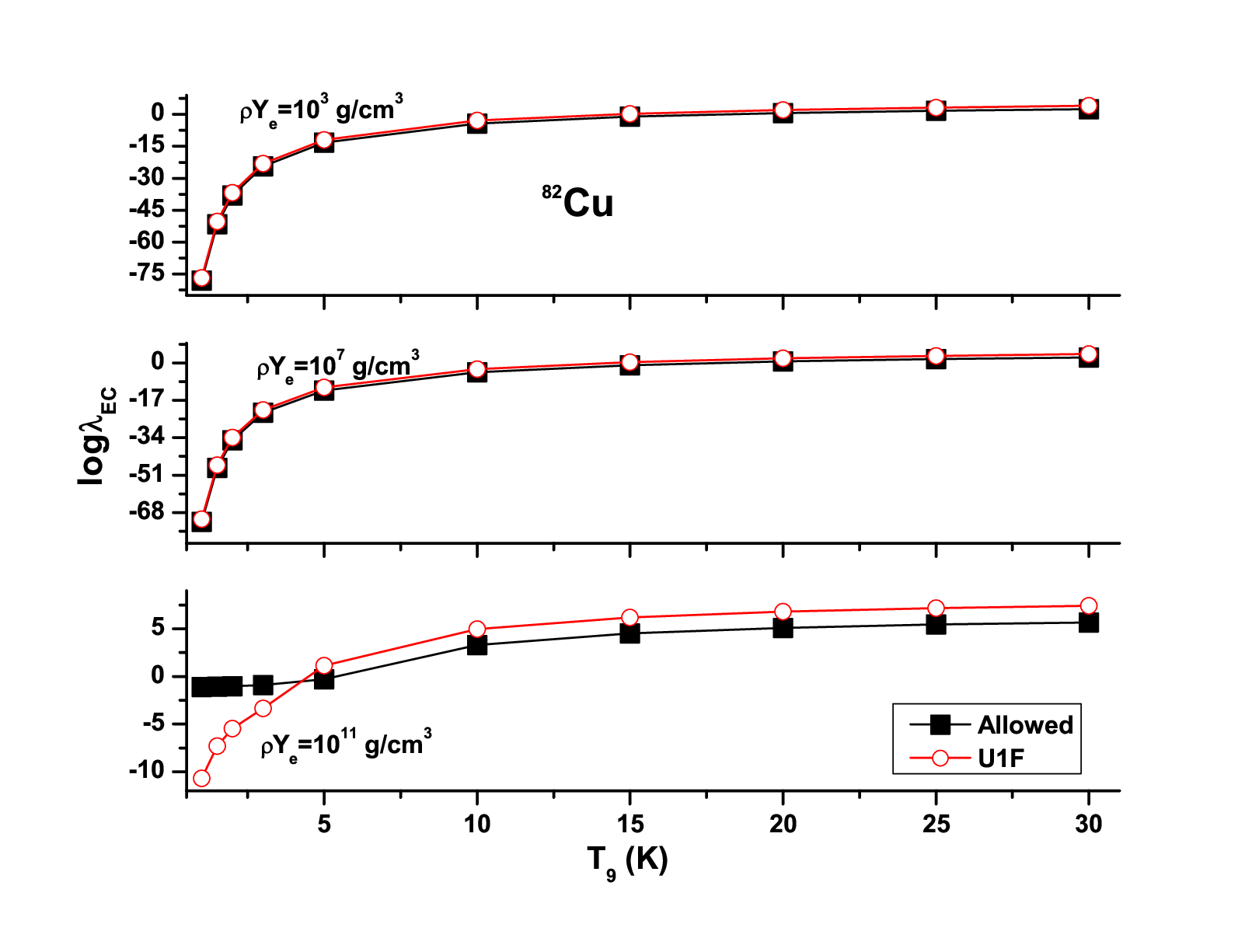}\\
\end{tabular}

\caption{(Color online) Calculated allowed GT and U1F electron
capture ($\lambda_{EC}$) rates on $^{73,74,77,78,81,82}$Cu in
stellar matter as a function of stellar temperature ($T_{9}$) at
selected stellar densities. The calculated capture rates are
tabulated in log to base 10 scale in units of s$^{-1}$.}
\label{figure9}
\end{center}
\end{figure*}

We calculate the EC ($\lambda_{EC}$) and PC ($\lambda_{PC}$) rates,
on copper isotopes, for both allowed GT and U1F transitions, for a
broad range of stellar temperature ($0.01 \times 10^{9} \leq T(K)
\leq 30 \times 10^{9})$ and density  $(10 \leq \rho Y_{e} (g
cm^{-3}) \leq 10^{11})$. Fig. \ref{figure9} depicts the calculated
EC rates on selected neutron-rich copper isotopes as a function of
stellar temperature in units of T$_9$ (which represent the core
temperature in units of 10$^{9}$K). The calculated capture rates are
shown at three different stellar density values of 10$^{3}$
gcm$^{-3}$ (depicting low density regions), 10$^{7}$ gcm$^{-3}$
(intermediate density regions) and 10$^{10}$ gcm$^{-3}$ (high
density regions). The pn-QRPA calculated capture rates are given in
logarithmic (to base 10) scales. We observe that the calculated EC
rates, both allowed GT and U1F, increase as the stellar temperature
and core density rise. Fig. \ref{figure9} clearly show that the U1F
capture weak-rates compete well with allowed GT capture rates and
the two rates have orders of magnitude differences.

\begin{figure*}[h]
\begin{center}
\begin{tabular}{cc}
\centering
\includegraphics[width=0.47\textwidth]{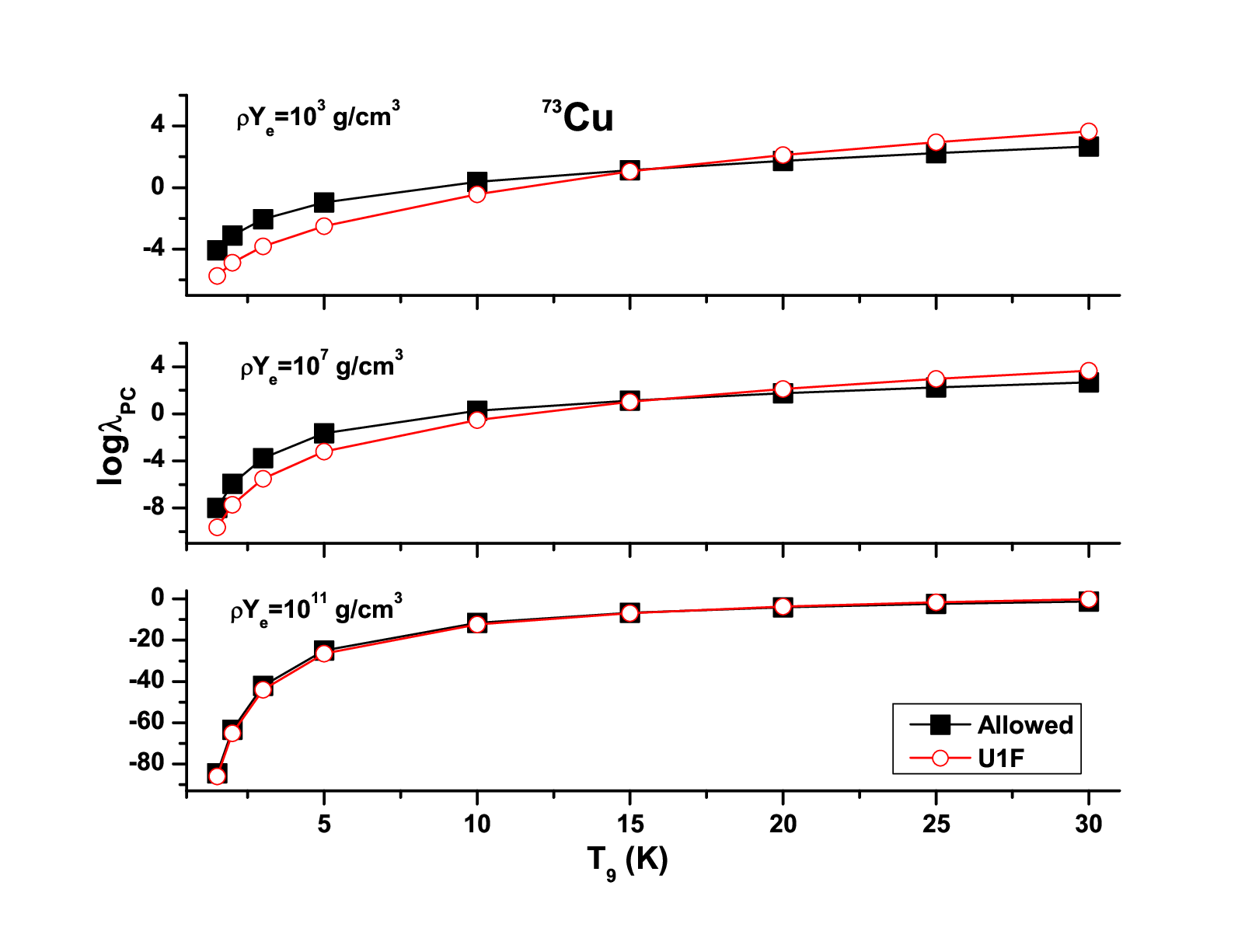}&
\includegraphics[width=0.47\textwidth]{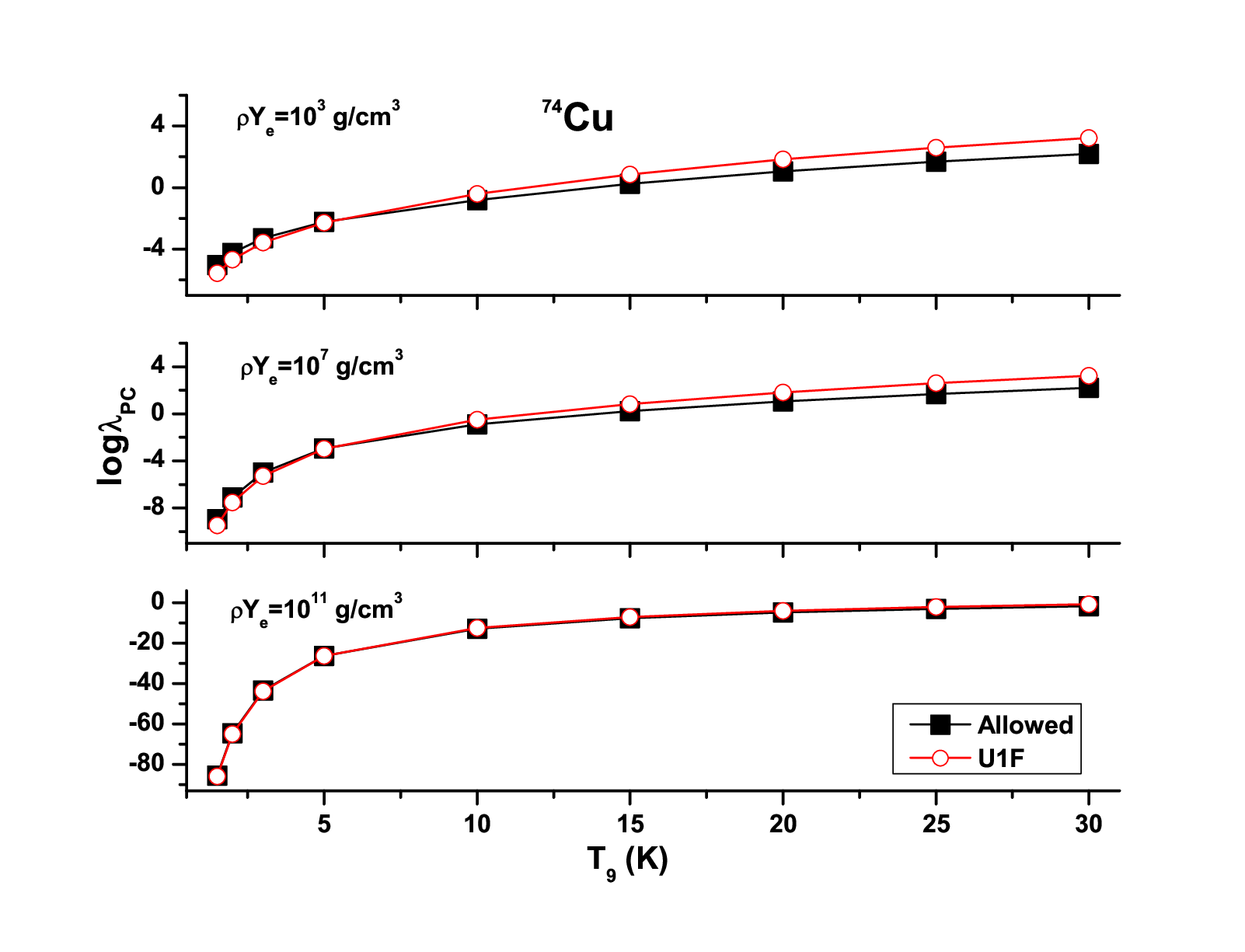}\\
\includegraphics[width=0.47\textwidth]{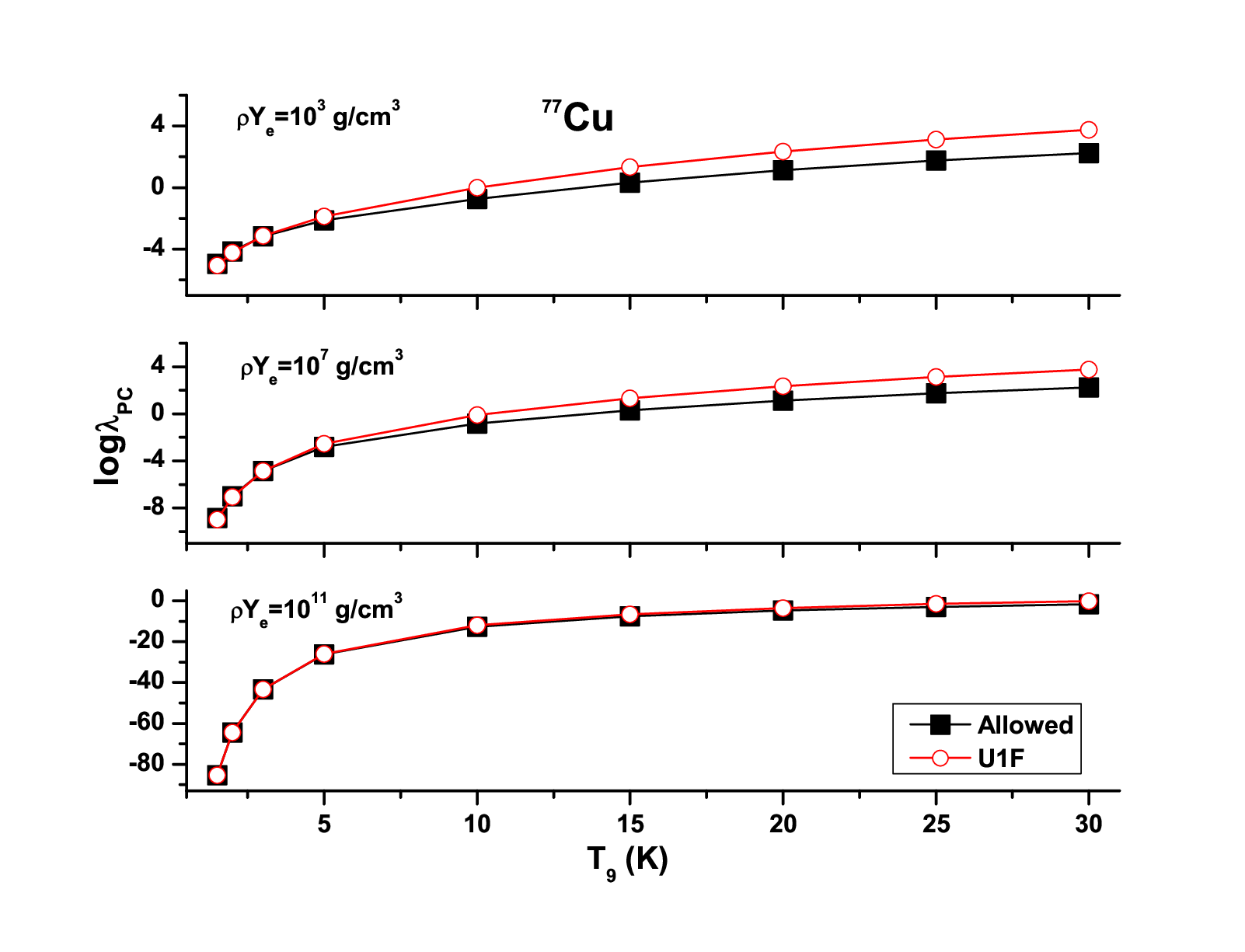}&
\includegraphics[width=0.47\textwidth]{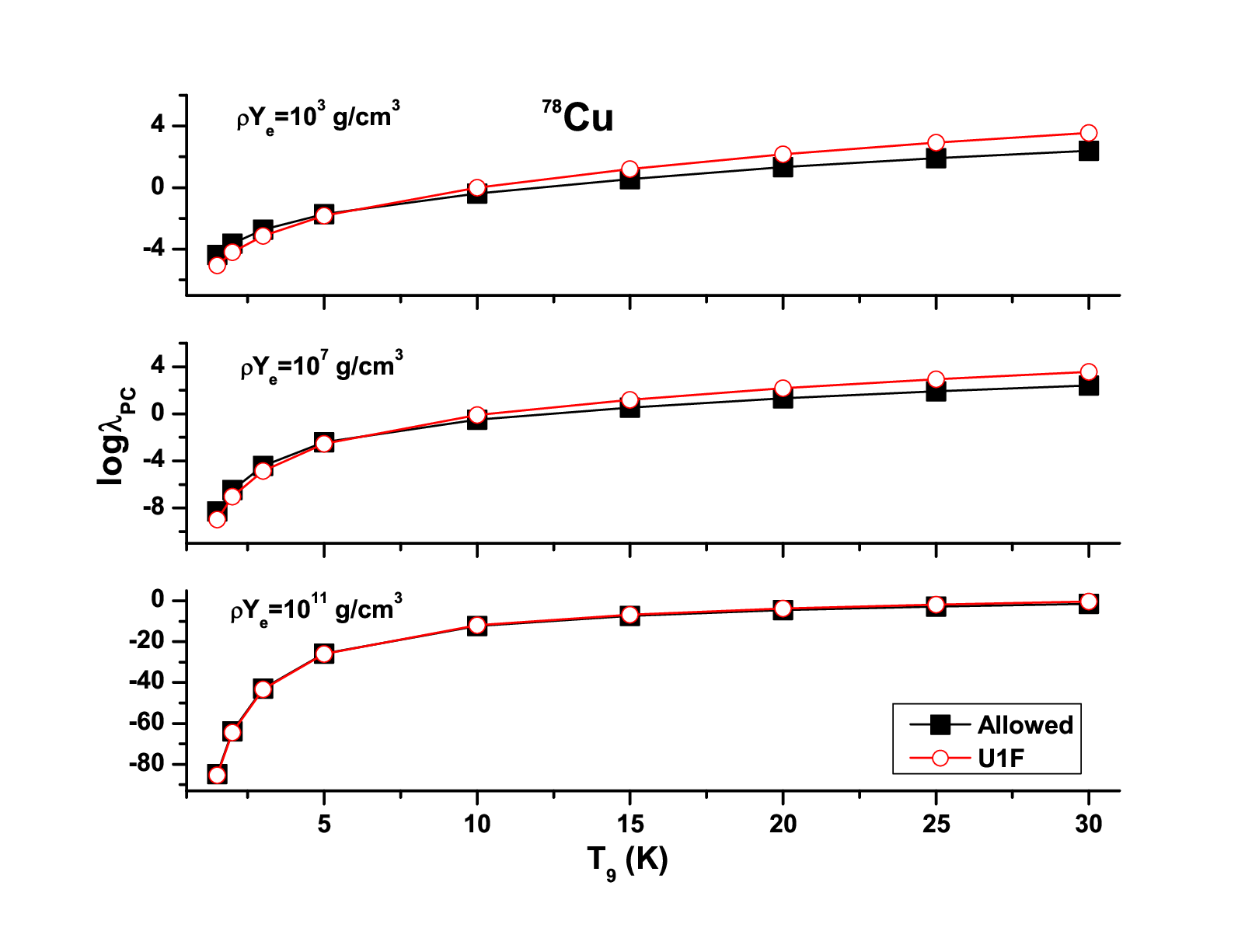}\\
\includegraphics[width=0.47\textwidth]{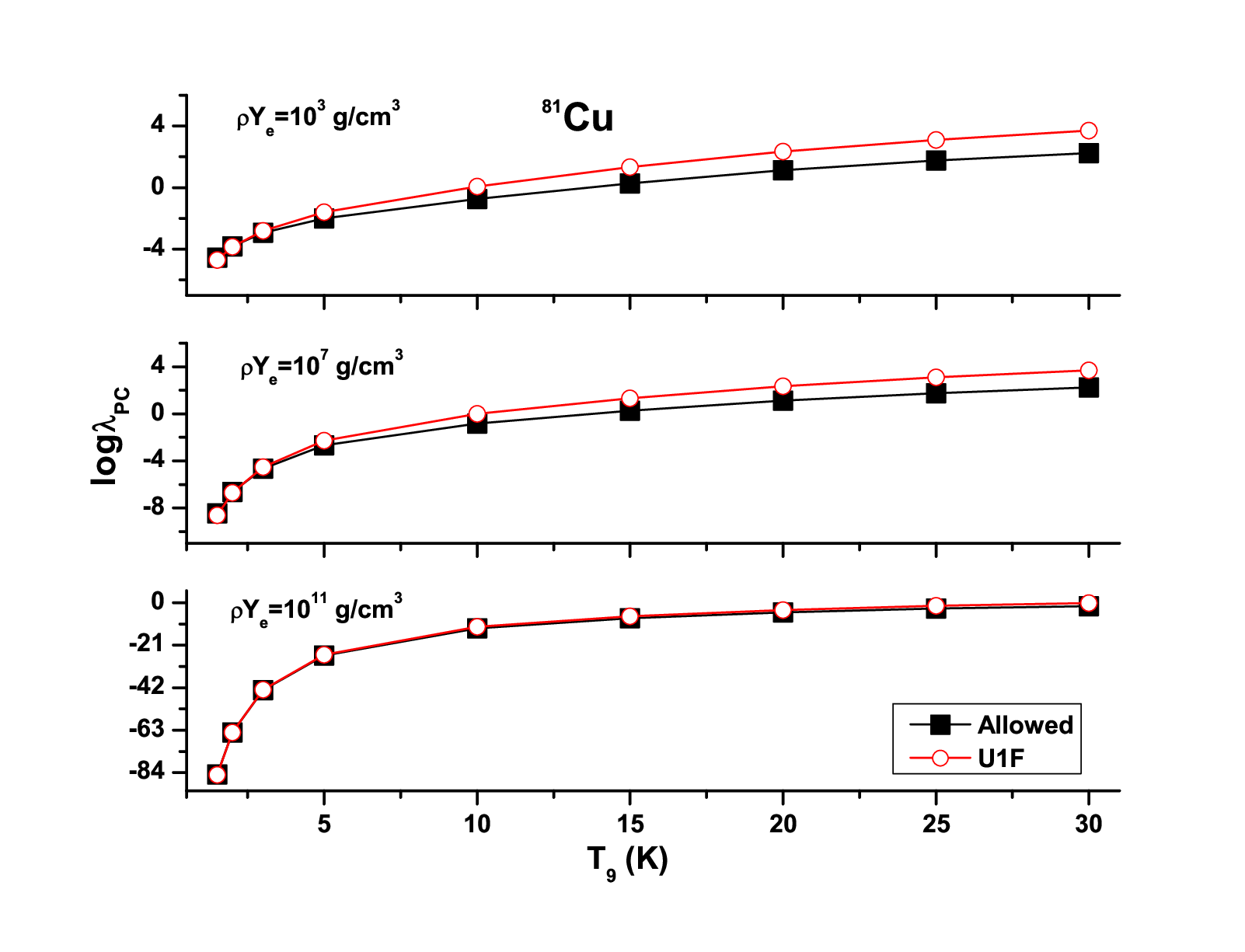}&
\includegraphics[width=0.47\textwidth]{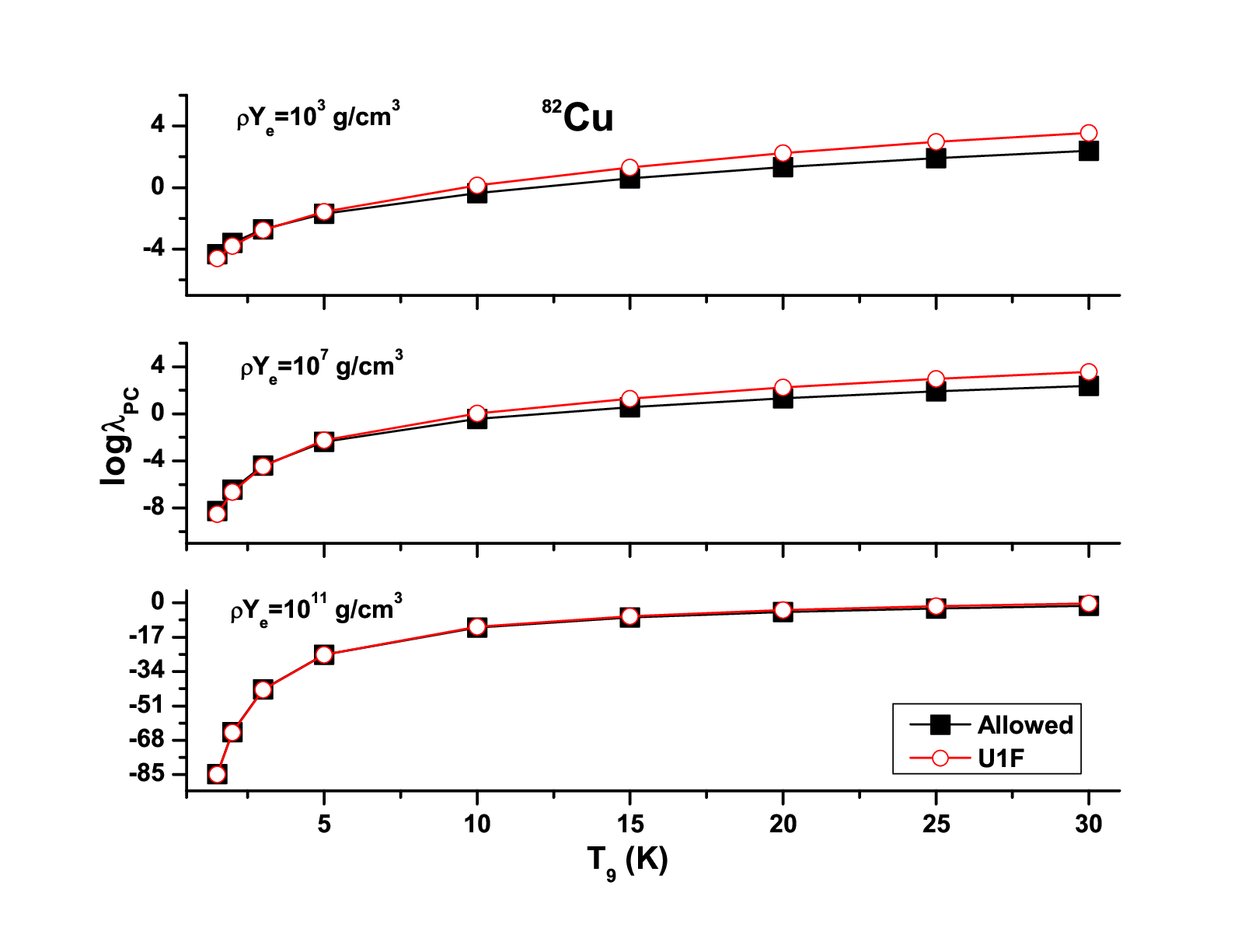}\\
\end{tabular}
\caption{(Color online) Same as Fig.~\ref{figure9} but for
calculated positron capture rates ($\lambda_{PC}$).}
\label{figure10}
\end{center}
\end{figure*}

Fig. \ref{figure10} shows similar result for pn-QRPA calculated PC
rates on selected copper isotopes. Here one notes that the allowed
GT and U1F rates are almost same and differ mostly at high stellar
temperatures. Tables~\ref{Table 2},~\ref{Table 3} and ~\ref{Table 4}
show the calculated EC and PC rates for $^{72,73,75,76,79,80}$Cu
isotopes at selected temperature and density values.  Once again all
calculated rates are given in log to base 10 scales. Complete ASCII
files of lepton capture rates, suitable for interpolation purposes
and use in simulation codes, may be requested from the corresponding
author.

\begin{figure*}[h]
\centering
\includegraphics[scale=0.60]{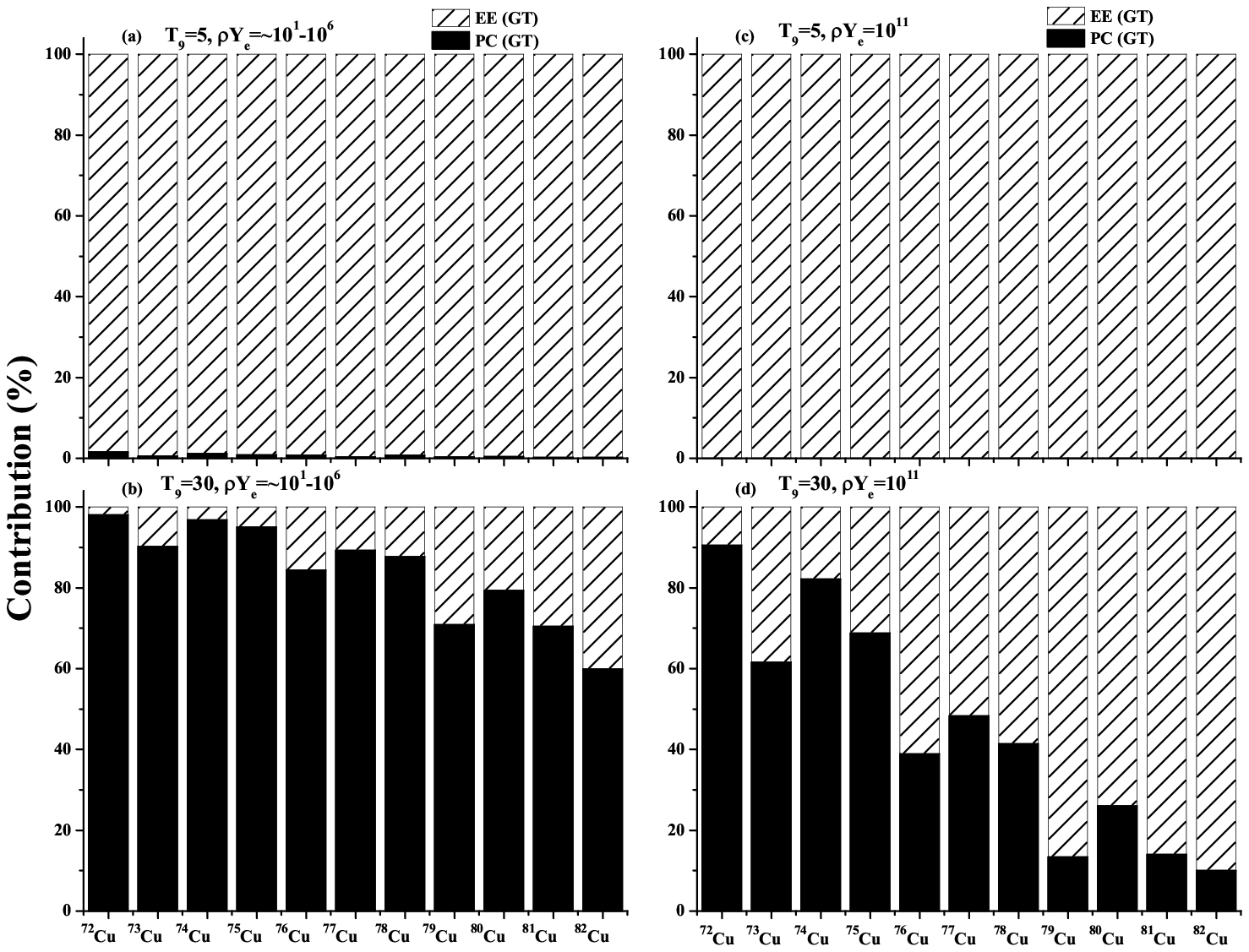}
\caption{Percentage contribution of allowed positron capture and
$\beta$-decay rates for neutron-rich copper isotopes. T$_{9}$ are
given in units of 10$^{9}$ K. Stellar density, $\rho$Y$_{e}$, is
given in units of g/cm$^{3}$.} \label{figure11}
\end{figure*}

\begin{figure*}[h]
\centering
\includegraphics[scale=0.60]{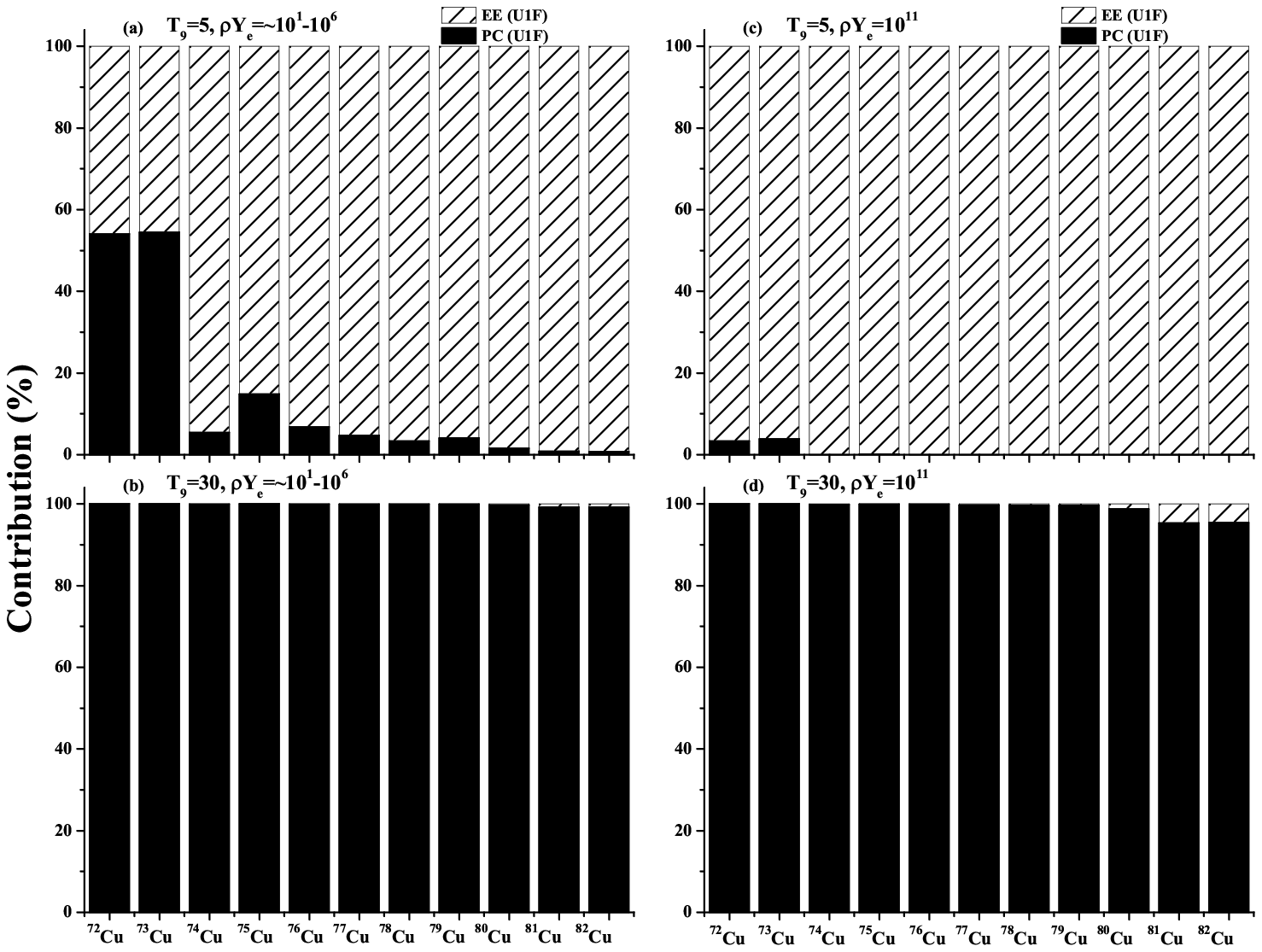}
\caption{Same as Fig.~\ref{figure11}, but for U1F rates.}
\label{figure12}
\end{figure*}

The ratio of electron to baryon (Y$_{e}$) increases as the electron
emission (EE) (presented in Ref. \cite{Maj17}) and PC rates
increase. One important investigation would be to find out  how the
two rates compete with each other for these neutron-rich copper
isotopes. In Fig.~\ref{figure11} and Fig.~\ref{figure12}, the
percentage contribution of EE and PC rates are shown. In
Fig.~\ref{figure11}, the allowed PC and EE rates are shown at T$_9$
= 5 (upper panels) and T$_9$ = 30 (lower panels). The left panels
show the situation at low-to-medium density while the right panels
depict the percentage contribution at high stellar density of
10$^{11}$gcm$^{-3}$. It is evident from Fig.~\ref{figure11} that PC
rates must be taken into consideration at high stellar temperatures
as they  dominate the competing EE rates for most of the copper
isotopes. Fig.~\ref{figure12} shows similar results for the U1F
rates. At T$_9$ = 30, the calculated PC rates contribute almost
100$\%$ for all copper isotopes. For $^{72,73}$Cu, even at T$_9$ = 5
and low-to-medium density regions, the PC contributes more than
50$\%$ to the total weak rates. These findings are crucial and
emphasize that $\it{both}$ EC and PC rates of copper isotopes need
to be taken into account in all prespernova evolution simulation
codes at high temperatures.

\section{Conclusions}

We used the density dependent RMF model to study the nuclear
ground-state properties of neutron-rich copper isotopes. Two
different interactions, namely the DD-ME2 and DD-PC1 interactions,
were employed to calculate the binding energy, proton and neutron
radii, root mean square (rms) charge radius, deformation parameter
and quadrupole moment of neutron-rich copper isotopes
($^{71-82}$Cu). The predictions of RMF model with the DD-ME2
functional for $BE$ per nucleon of the isotopic chain were found in
agreement with experimental data and the results of FRDM. We
obtained PECs of $^{71-82}$Cu isotopes using the quadrupole moment
constrained RMF model calculation. These PECs indicate that
$^{71}$Cu and $^{79}$Cu nuclei have spherical shape in their
ground-state while shape of $^{78}$Cu and $^{80}$Cu nuclei are close
to spherical character. The shape of others are predicted as prolate
in our RMF calculation.

The deformed pn-QRPA theoretical model was used in the later half of
this work to calculate lepton capture weak-rates on copper nuclide
in stellar scenario. Here we used the pn-QRPA model to compute the
allowed GT and U1F transition strengths for these copper isotopes.
All ground-state and excited states charge-changing strength
distributions were computed in a microscopic way. Later we also
calculated the stellar EC and PC rates including both allowed GT and
U1F contributions. This completes our initial study where we
presented the $\beta$-decay rates of neutron-rich copper nuclide in
stellar matter. The EC rates on copper isotopes were found to be
important specially in high temperature and high density regions. It
was concluded that at high stellar temperature the PC rates dominate
 the corresponding $\beta$-decay rates and should be taken into account by
core-collapse simulators to depict a realistic picture of the
process.

\vspace{0.5in} \textbf{Acknowledgment}:  J.-U. Nabi would like to
acknowledge the support of the Higher Education Commission Pakistan
through project number 5557/KPK/NRPU/R$\&$D/HEC/2016 and the
Pakistan Science Foundation through project number
PSF-TUBITAK/KP-GIKI (02). This research has been supported by the
Council of Higher Education of Turkey (Mevlana Exchange Program
Based on Project) Project Number: MEV-2016-094, 2016.

\newpage

\clearpage
\begin{table*}[h]
\centering \scriptsize \caption{The RMF calculated ground-state properties of $^{71-82}$Cu nuclei using the density dependent DD-ME2 interaction} {%
\begin{tabular}{@{}ccccccc}
{Nuclei} & $BE/A$ (MeV) & $r_{n}$ (fm) & $r_{p}$ (fm) &  $r_{c}$ (fm) & $\beta_{2}$ & $Q_{T}$ (barn) \\
\hline
$^{71}\textrm{Cu}$\hphantom{00} & \hphantom{0}8.634& \hphantom{0}4.048& \hphantom{0}4.853& \hphantom{0}3.935 & \hphantom{0}0.000& \hphantom{0}0.000\\
$^{72}\textrm{Cu}$\hphantom{00} & \hphantom{0}8.601& \hphantom{0}4.071& \hphantom{0}4.861& \hphantom{0}3.943 & \hphantom{0}0.110& \hphantom{0}1.499\\
$^{73}\textrm{Cu}$\hphantom{00} & \hphantom{0}8.569& \hphantom{0}4.107& \hphantom{0}4.879& \hphantom{0}3.961 & \hphantom{0}0.113& \hphantom{0}1.575\\
$^{74}\textrm{Cu}$\hphantom{00} & \hphantom{0}8.534& \hphantom{0}4.132& \hphantom{0}4.890& \hphantom{0}3.972 & \hphantom{0}0.115& \hphantom{0}1.640\\
$^{75}\textrm{Cu}$\hphantom{00} & \hphantom{0}8.494& \hphantom{0}4.158& \hphantom{0}4.903& \hphantom{0}3.984 & \hphantom{0}0.116& \hphantom{0}1.685\\
$^{76}\textrm{Cu}$\hphantom{00} & \hphantom{0}8.447& \hphantom{0}4.166& \hphantom{0}4.901& \hphantom{0}3.982 & \hphantom{0}0.115& \hphantom{0}1.703\\
$^{77}\textrm{Cu}$\hphantom{00} & \hphantom{0}8.399& \hphantom{0}4.181& \hphantom{0}4.906& \hphantom{0}3.987 & \hphantom{0}0.111& \hphantom{0}1.688\\
$^{78}\textrm{Cu}$\hphantom{00} & \hphantom{0}8.351& \hphantom{0}4.191& \hphantom{0}4.907& \hphantom{0}3.988 & \hphantom{0}0.060& \hphantom{0}0.294\\
$^{79}\textrm{Cu}$\hphantom{00} & \hphantom{0}8.310& \hphantom{0}4.204& \hphantom{0}4.912& \hphantom{0}3.993 & \hphantom{0}0.000& \hphantom{0}0.000\\
$^{80}\textrm{Cu}$\hphantom{00} & \hphantom{0}8.229& \hphantom{0}4.251& \hphantom{0}4.921& \hphantom{0}4.002 & \hphantom{0}0.020& \hphantom{0}0.000\\
$^{81}\textrm{Cu}$\hphantom{00} & \hphantom{0}8.152& \hphantom{0}4.302& \hphantom{0}4.939& \hphantom{0}4.019 & \hphantom{0}0.109& \hphantom{0}1.805\\
$^{82}\textrm{Cu}$\hphantom{00} & \hphantom{0}8.076& \hphantom{0}4.345& \hphantom{0}4.952& \hphantom{0}4.032 & \hphantom{0}0.117& \hphantom{0}1.969\\

\end{tabular}%
\label{Table 1} }
\end{table*}

\begin{table*}
\centering \scriptsize \caption{Calculated allowed (GT) and unique
first-forbidden (U1F) lepton capture rates on $^{72,73}$Cu for
different selected densities and temperatures in stellar matter. The
first column shows the stellar density ($\rho$Y$_{e}$) (in units of
gcm$^{-3}$). T$_{9}$ are given in units of 10$^{9}$ K. The
calculated capture rates are tabulated in logarithmic (to base 10)
scale in units of s$^{-1}$}\label{Table 2}
\begin{tabular}{|c|c|cccc|cccc|}

$\rho$$\it Y_{e}$ & T$_{9}$ & \multicolumn{4}{c|}{$^{72}$Cu}& \multicolumn{4}{c|}{$^{73}$Cu}\\
\cline{3-10} & &{$\lambda_{PC}$ (GT)} & {$\lambda_{EC}$ (GT)} & {$\lambda_{PC}$ (U1F)} & {$\lambda_{EC}$ (U1F)}&{$\lambda_{PC}$ (GT)} & {$\lambda_{EC}$ (GT)} & {$\lambda_{PC}$ (U1F)} & {$\lambda_{EC}$ (U1F)}\\
\hline
      & 1.5   & -5.46 & -30.00 & -5.47 & -28.75 & -4.05 & -33.03 & -5.72 & -32.19 \\
      & 2     & -4.68 & -22.44 & -4.58 & -21.19 & -3.11 & -24.71 & -4.88 & -23.71 \\
      & 3     & -3.71 & -14.69 & -3.44 & -13.42 & -2.05 & -16.20 & -3.81 & -15.06 \\
      & 5     & -2.63 & -8.18 & -2.08 & -6.88 & -0.96 & -9.10 & -2.51 & -7.88 \\
 10$^{2}$     & 10    & -1.15 & -2.74 & -0.09 & -1.37 & 0.36  & -3.04 & -0.44 & -1.70 \\
     & 15    & -0.09 & -0.48 & 1.27  & 0.96  & 1.14  & -0.55 & 1.04  & 0.91 \\
     & 20    & 0.74  & 0.90  & 2.28  & 2.39  & 1.74  & 0.89  & 2.12  & 2.44 \\
     & 25    & 1.39  & 1.85  & 3.07  & 3.40  & 2.24  & 1.86  & 2.96  & 3.47 \\
      & 30    & 1.90  & 2.55  & 3.73  & 4.15  & 2.66  & 2.57  & 3.66  & 4.23 \\
\hline
      & 1.5   & -7.48 & -27.98 & -7.49 & -26.73 & -6.07 & -31.00 & -7.74 & -30.16 \\
      & 2     & -5.96 & -21.15 & -5.87 & -19.89 & -4.40 & -23.41 & -6.17 & -22.42 \\
      & 3     & -4.21 & -14.18 & -3.94 & -12.91 & -2.55 & -15.70 & -4.31 & -14.56 \\
      & 5     & -2.72 & -8.08 & -2.17 & -6.79 & -1.05 & -9.01 & -2.60 & -7.79 \\
 10$^{6}$     & 10    & -1.16 & -2.73 & -0.10 & -1.36 & 0.35  & -3.03 & -0.45 & -1.69 \\
      & 15    & -0.09 & -0.47 & 1.26  & 0.96  & 1.14  & -0.55 & 1.04  & 0.91 \\
     & 20    & 0.74  & 0.90  & 2.28  & 2.40  & 1.74  & 0.89  & 2.12  & 2.44 \\
      & 25    & 1.39  & 1.85  & 3.07  & 3.40  & 2.24  & 1.86  & 2.96  & 3.47 \\
      & 30    & 1.90  & 2.55  & 3.73  & 4.15  & 2.66  & 2.57  & 3.66  & 4.23 \\
\hline
      & 1.5   & -42.80 & 1.82  & -42.80 & 3.15  & -41.39 & 0.12  & -43.06 & -0.42 \\
    & 2     & -32.67 & 1.87  & -32.57 & 3.20  & -31.10 & 0.42  & -32.87 & 0.59 \\
     & 3     & -22.35 & 1.96  & -22.08 & 3.30  & -20.69 & 0.83  & -22.45 & 1.60 \\
     & 5     & -13.76 & 2.12  & -13.22 & 3.47  & -12.09 & 1.35  & -13.65 & 2.45 \\
 10$^{10}$    & 10    & -6.62 & 2.52  & -5.57 & 3.91  & -5.11 & 2.23  & -5.92 & 3.58 \\
     & 15    & -3.64 & 2.98  & -2.30 & 4.43  & -2.40 & 2.90  & -2.52 & 4.38 \\
     & 20    & -1.82 & 3.41  & -0.30 & 4.92  & -0.82 & 3.40  & -0.46 & 4.96 \\
     & 25    & -0.57 & 3.77  & 1.11  & 5.33  & 0.29  & 3.78  & 0.99  & 5.40 \\
    & 30    & 0.37  & 4.06  & 2.20  & 5.67  & 1.14  & 4.08  & 2.13  & 5.75 \\
\hline

\end{tabular}
\end{table*}

\begin{table*}
\centering \scriptsize \caption{Same as Table~\ref{Table 2} but for
$^{75,76}$Cu}\label{Table 3}

\begin{tabular}{|c|c|cccc|cccc|}

$\rho$$\it Y_{e}$ & T$_{9}$ & \multicolumn{4}{c|}{$^{75}$Cu}& \multicolumn{4}{c|}{$^{76}$Cu}\\
\cline{3-10} & &{$\lambda_{PC}$ (GT)} & {$\lambda_{EC}$ (GT)} & {$\lambda_{PC}$ (U1F)} & {$\lambda_{EC}$ (U1F)}&{$\lambda_{PC}$ (GT)} & {$\lambda_{EC}$ (GT)} & {$\lambda_{PC}$ (U1F)} & {$\lambda_{EC}$ (U1F)}\\
\hline
     & 1.5   & -5.01 & -38.87 & -5.36 & -37.96 & -4.24 & -34.30 & -5.15 & -33.06 \\
      & 2     & -4.27 & -28.96 & -4.53 & -27.88 & -3.42 & -25.54 & -4.30 & -24.29 \\
      & 3     & -3.39 & -18.81 & -3.48 & -17.58 & -2.43 & -16.61 & -3.22 & -15.35 \\
      & 5     & -2.41 & -10.33 & -2.19 & -8.98 & -1.35 & -9.20 & -1.91 & -7.91 \\
 10$^{2}$     & 10    & -1.01 & -3.36 & -0.20 & -1.89 & 0.04  & -3.05 & -0.05 & -1.68 \\
     & 15    & 0.04  & -0.68 & 1.24  & 0.88  & 1.01  & -0.53 & 1.22  & 0.89 \\
      & 20    & 0.87  & 0.82  & 2.29  & 2.45  & 1.77  & 0.94  & 2.22  & 2.42 \\
      & 25    & 1.51  & 1.82  & 3.08  & 3.50  & 2.35  & 1.92  & 3.00  & 3.47 \\
      & 30    & 2.03  & 2.55  & 3.72  & 4.27  & 2.82  & 2.64  & 3.65  & 4.24 \\
\hline
      & 1.5   & -7.03 & -36.84 & -7.39 & -35.93 & -6.26 & -32.28 & -7.18 & -31.04 \\
      & 2     & -5.56 & -27.67 & -5.82 & -26.59 & -4.71 & -24.24 & -5.59 & -23.00 \\
      & 3     & -3.89 & -18.31 & -3.98 & -17.07 & -2.92 & -16.10 & -3.72 & -14.84 \\
      & 5     & -2.50 & -10.24 & -2.28 & -8.89 & -1.43 & -9.11 & -2.00 & -7.82 \\
 10$^{6}$     & 10    & -1.02 & -3.35 & -0.21 & -1.88 & 0.03  & -3.04 & -0.06 & -1.67 \\
     & 15    & 0.03  & -0.68 & 1.24  & 0.88  & 1.00  & -0.53 & 1.22  & 0.90 \\
      & 20    & 0.87  & 0.82  & 2.29  & 2.45  & 1.77  & 0.94  & 2.22  & 2.43 \\
      & 25    & 1.51  & 1.83  & 3.08  & 3.50  & 2.35  & 1.92  & 3.00  & 3.47 \\
      & 30    & 2.03  & 2.55  & 3.72  & 4.27  & 2.82  & 2.64  & 3.65  & 4.24 \\
\hline
     & 1.5   & -42.34 & -6.89 & -42.70 & -6.43 & -41.57 & -2.98 & -42.49 & -1.66 \\
    & 2     & -32.27 & -4.89 & -32.52 & -3.75 & -31.41 & -1.58 & -32.30 & -0.25 \\
     & 3     & -22.03 & -2.32 & -22.12 & -1.01 & -21.06 & -0.16 & -21.86 & 1.17 \\
     & 5     & -13.54 & -0.11 & -13.33 & 1.31  & -12.48 & 1.01  & -13.06 & 2.36 \\
 10$^{10}$    & 10    & -6.49 & 1.88  & -5.68 & 3.38  & -5.43 & 2.18  & -5.53 & 3.58 \\
     & 15    & -3.52 & 2.76  & -2.32 & 4.35  & -2.54 & 2.91  & -2.34 & 4.36 \\
     & 20    & -1.70 & 3.33  & -0.29 & 4.97  & -0.79 & 3.44  & -0.36 & 4.95 \\
     & 25    & -0.44 & 3.74  & 1.12  & 5.43  & 0.41  & 3.84  & 1.04  & 5.40 \\
    & 30    & 0.51  & 4.06  & 2.19  & 5.79  & 1.30  & 4.15  & 2.12  & 5.75 \\

\hline

\end{tabular}
\end{table*}

\begin{table*}
\centering \scriptsize \caption{Same as Table~\ref{Table 2} but for
$^{79,80}$Cu }\label{Table 4}
    \begin{tabular}{|c|c|cccc|cccc|}

$\rho$$\it Y_{e}$ & T$_{9}$ & \multicolumn{4}{c|}{$^{79}$Cu}& \multicolumn{4}{c|}{$^{80}$Cu}\\
\cline{3-10} & &{$\lambda_{PC}$ (GT)} & {$\lambda_{EC}$ (GT)} & {$\lambda_{PC}$ (U1F)} & {$\lambda_{EC}$ (U1F)}&{$\lambda_{PC}$ (GT)} & {$\lambda_{EC}$ (GT)} & {$\lambda_{PC}$ (U1F)} & {$\lambda_{EC}$ (U1F)}\\
\hline
     & 1.5   & -4.36 & -49.99 & -4.97 & -49.01 & -4.37 & -45.64 & -4.76 & -44.41 \\
      & 2     & -3.59 & -36.98 & -4.12 & -35.91 & -3.62 & -33.81 & -3.93 & -32.57 \\
      & 3     & -2.67 & -23.78 & -3.05 & -22.61 & -2.72 & -21.79 & -2.88 & -20.54 \\
      & 5     & -1.71 & -12.90 & -1.78 & -11.63 & -1.75 & -11.88 & -1.64 & -10.60 \\
10$^{2}$      & 10    & -0.42 & -4.19 & -0.04 & -2.80 & -0.48 & -3.89 & 0.10  & -2.53 \\
      & 15    & 0.65  & -0.98 & 1.25  & 0.47  & 0.44  & -0.90 & 1.27  & 0.53 \\
      & 20    & 1.48  & 0.75  & 2.28  & 2.24  & 1.19  & 0.73  & 2.21  & 2.22 \\
      & 25    & 2.10  & 1.85  & 3.07  & 3.40  & 1.78  & 1.79  & 2.95  & 3.33 \\
      & 30    & 2.58  & 2.64  & 3.70  & 4.23  & 2.25  & 2.54  & 3.56  & 4.14 \\
\hline
     & 1.5   & -6.39 & -47.96 & -7.00 & -46.99 & -6.40 & -43.61 & -6.78 & -42.38 \\
      & 2     & -4.87 & -35.68 & -5.41 & -34.62 & -4.91 & -32.52 & -5.22 & -31.28 \\
      & 3     & -3.16 & -23.27 & -3.54 & -22.11 & -3.22 & -21.29 & -3.38 & -20.04 \\
      & 5     & -1.80 & -12.81 & -1.87 & -11.54 & -1.84 & -11.79 & -1.73 & -10.51 \\
 10$^{6}$     & 10    & -0.43 & -4.18 & -0.05 & -2.79 & -0.49 & -3.88 & 0.09  & -2.52 \\
      & 15    & 0.65  & -0.98 & 1.24  & 0.47  & 0.43  & -0.90 & 1.27  & 0.53 \\
      & 20    & 1.48  & 0.75  & 2.28  & 2.24  & 1.18  & 0.73  & 2.21  & 2.22 \\
      & 25    & 2.10  & 1.85  & 3.07  & 3.40  & 1.77  & 1.79  & 2.95  & 3.33 \\
      & 30    & 2.58  & 2.64  & 3.70  & 4.23  & 2.25  & 2.54  & 3.56  & 4.14 \\
\hline
     & 1.5   & -41.70 & -15.97 & -42.31 & -17.95 & -41.71 & -14.79 & -42.10 & -13.46 \\
     & 2     & -31.58 & -12.07 & -32.12 & -12.12 & -31.61 & -10.18 & -31.92 & -8.85 \\
     & 3     & -21.30 & -7.39 & -21.69 & -6.24 & -21.35 & -5.54 & -21.52 & -4.21 \\
     & 5     & -12.84 & -2.76 & -12.92 & -1.43 & -12.88 & -1.77 & -12.78 & -0.42 \\
 10$^{10}$    & 10    & -5.88 & 1.02  & -5.52 & 2.45  & -5.94 & 1.32  & -5.39 & 2.71 \\
     & 15    & -2.89 & 2.46  & -2.32 & 3.93  & -3.11 & 2.54  & -2.29 & 3.99 \\
     & 20    & -1.07 & 3.25  & -0.30 & 4.76  & -1.37 & 3.23  & -0.37 & 4.74 \\
     & 25    & 0.16  & 3.77  & 1.11  & 5.33  & -0.17 & 3.70  & 0.99  & 5.26 \\
     & 30    & 1.06  & 4.14  & 2.16  & 5.74  & 0.73  & 4.04  & 2.02  & 5.65 \\

\hline

\end{tabular}
\end{table*}

\clearpage
\newpage
%

%
%

\end{document}